\documentclass[12pt]{article}%
\usepackage{etex}
\usepackage{caption}

\usepackage{lscape}

\usepackage{xy}
\input xy
\xyoption{all}
\usepackage[nosort]{cite}
\usepackage{graphicx}
\usepackage{multicol}
\usepackage{amsfonts}
\usepackage{amssymb}
\usepackage{amsmath}
\usepackage{dsfont}
\usepackage{heck}
\usepackage{afterpage}
\usepackage{setspace}
\usepackage{verbatim}
\usepackage{color}
\usepackage{longtable}
\usepackage{float}
\usepackage{subcaption}
\usepackage{epsfig}
\usepackage{epstopdf}
\usepackage{adjustbox}
\usepackage{todonotes}
\usepackage[makeroom]{cancel}
\usepackage{tikz}
\usepackage[margin=1in]{geometry}
\usepackage{titletoc}
\usepackage{hyperref}%

\usepackage{multirow}

\providecommand{\U}[1]{\protect\rule{.1in}{.1in}}
\pdfoutput=1
\newsavebox{\mysavebox}

\usetikzlibrary[shapes.geometric]
\usetikzlibrary{positioning}
\usetikzlibrary{calc,intersections,through,backgrounds}
\tikzset{>=stealth}
\usetikzlibrary{decorations.pathmorphing}
\usetikzlibrary{decorations.markings}
\tikzset{>=stealth}
\hypersetup{colorlinks,citecolor=black,filecolor=black,linkcolor=black,urlcolor=black,pdftex}
\usetikzlibrary{decorations.markings}

\numberwithin{equation}{section}

\newcommand{\ba}{\begin{eqnarray}}
\newcommand{\ea}{\end{eqnarray}}

\newcommand{\be}{\begin{equation}}
\newcommand{\ee}{\end{equation}}

\tikzstyle{startstop} = [rectangle, rounded corners, minimum width=3cm, minimum height=1cm,text centered, draw=black, fill=blue!10]
\tikzstyle{startstop} = [rectangle, rounded corners, minimum width=3cm, minimum height=1cm,text centered, draw=black, fill=blue!10]
\tikzstyle{io} = [trapezium, trapezium left angle=70, trapezium right angle=110, minimum width=3cm, minimum height=1cm, text centered, draw=black, fill=blue!30]
\tikzstyle{process} = [rectangle, minimum width=3cm, minimum height=1cm, text centered, draw=black, fill=orange!30]
\tikzstyle{decision} = [diamond, minimum width=3cm, minimum height=1cm, text centered, draw=black, fill=green!30]
\tikzstyle{arrow} = [thick,->,>=stealth]
\begin{document}

\date{March 2017}

\title{6D SCFTs and Phases of 5D Theories}

\institution{SCGP}{\centerline{${}^{1}$Simons Center for Geometry and Physics, Stony Brook, NY 11794, USA}}

\institution{UNC}{\centerline{${}^{2}$Department of Physics, University of North Carolina, Chapel Hill, NC 27599, USA}}

\institution{UCSBmath}{\centerline{${}^{3}$Department of Mathematics, University of California Santa Barbara, CA 93106, USA}}

\institution{UCSBphys}{\centerline{${}^{4}$Department of Physics, University of California Santa Barbara, CA 93106, USA}}

\authors{Michele Del Zotto\worksat{\SCGP}\footnote{e-mail: {\tt mdelzotto@scgp.stonybrook.edu}},
Jonathan J. Heckman\worksat{\UNC}\footnote{e-mail: {\tt jheckman@email.unc.edu}}, and
David R. Morrison\worksat{\UCSBmath, \UCSBphys}\footnote{e-mail: {\tt drm@physics.ucsb.edu}}}

\abstract{Starting from 6D superconformal field theories (SCFTs) realized via F-theory,
we show how reduction on a circle
leads to a uniform perspective on the phase structure of the resulting 5D theories, and their
possible conformal fixed points. Using the correspondence between
F-theory reduced on a circle and M-theory on the corresponding
elliptically fibered Calabi--Yau threefold,
we show that each 6D SCFT with minimal supersymmetry directly reduces to a
collection of between one and four 5D SCFTs. Additionally, we find
that in most cases, reduction of the tensor branch of a 6D SCFT yields a 5D generalization
of a quiver gauge theory. These two reductions of the theory often correspond
to different phases in the 5D theory which are in general connected by a sequence of
flop transitions in the extended K\"ahler cone of the Calabi--Yau threefold. We also elaborate on the
structure of the resulting conformal fixed points, and emergent flavor symmetries,
as realized by M-theory on a canonical singularity.}

\maketitle

\tableofcontents

\enlargethispage{\baselineskip}

\setcounter{tocdepth}{2}

\newpage

\section{Introduction \label{sec:INTRO}}

Developing tools to characterize interacting SCFTs in higher
spacetime dimensions is one of the challenges of
contemporary theoretical physics. These systems exhibit striking departures from the standard paradigm of lower dimensional
examples. The traditional methods of perturbation theory do
not apply, and one must instead
resort to stringy constructions to even establish existence.
One of the remarkable recent developments in string theory
is that not only do such theories exist, but many of their properties
can be understood by using the geometry of extra
dimensions.

Celebrated examples of this type are 6D superconformal field
theories (SCFTs) \cite{Witten:1995zh, Strominger:1995ac, Seiberg:1996qx}.
For theories with $(2,0)$ supersymmetry, there is an
ADE\ classification given by Type IIB\ on supersymmetric
orbifolds $\mathbb{C}^{2}/\Gamma_{\text{ADE}}$ (see also \cite{Seiberg:1997ax, Henningson:2004dh, Cordova:2015vwa}). For theories with $(1,0)$
supersymmetry, there is a related classification of the theories
which can be obtained from
F-theory \cite{Heckman:2013pva, Gaiotto:2014lca, DelZotto:2014hpa,
Heckman:2014qba, DelZotto:2014fia, Heckman:2015bfa, Bhardwaj:2015xxa}. Several features of these models are captured
by the above string constructions, for instance the moduli spaces of vacua are
captured by deformations of the Calabi--Yau geometry, the anomaly polynomials are encoded in the intersection theory of the F-theory base \cite{Sadov:1996zm,Grassi:2000wer,Grassi:2011hq}, and the 6D
omega-background partition function is captured by topological string amplitudes
 on the Calabi--Yau (see e.g. \cite{Lockhart:2012vp,Haghighat:2014vxa,Kim:2016foj,DelZotto:2016pvm,Gu:2017ccq}).

Compactification also yields insight into strongly coupled phases of lower-dimensional systems. For
example, in the case of the 6D~theories with $(2,0)$ supersymmetry, the
higher-dimensional perspective provides a geometric origin for non-trivial 4D
dualities \cite{Vafa:1997mh, Witten:1997sc, Argyres:2007cn, Gaiotto:2009we}.
Though there is reduced supersymmetry in the case of the 6D $(1,0)$ theories,
there has recently been significant progress in developing analogous results
\cite{DelZotto:2015isa, Gaiotto:2015usa, Ohmori:2015pua, Franco:2015jna, DelZotto:2015rca,
Hanany:2015pfa, Aganagic:2015cta, Ohmori:2015pia, Coman:2015bqq, Morrison:2016nrt, Heckman:2016xdl, Apruzzi:2016nfr, Razamat:2016dpl}.

Our aim in this work will be to use this 6D perspective to shed light on the
phase structure of 5D field theories. For earlier work on the construction
and study of such theories, see for example, \cite{Witten:1996qb,Seiberg:1996bd, Morrison:1996xf, Intriligator:1997pq,Diaconescu:1998cn}, and for
more recent studies, see for example \cite{Bergman:2012kr, Bergman:2013koa, Apruzzi:2014qva,Kim:2015jba,
Hayashi:2015fsa, Bergman:2015dpa, DHoker:2016ujz, DHoker:2016ysh}.
Stringy constructions of such 5D fixed points include
D-brane probes of singularities \cite{Douglas:1996sw}, suspended $(p,q)$ five-brane
webs \cite{Aharony:1997ju, Aharony:1997bh},
and purely geometric realizations using M-theory on a Calabi--Yau
threefold with a canonical singularity \cite{Witten:1996qb,Morrison:1996xf,Douglas:1996xp,Katz:1996fh,Intriligator:1997pq,Morrison:1998cs}.

One of the confusing issues in such 5D theories is the existence of rather tight
constraints on purely gauge theoretic constructions. Using only effective
field theory arguments, reference \cite{Intriligator:1997pq} argued that the strong coupling limit
of a 5D gauge theory can only produce a conformal fixed point when there is a
single simple gauge group factor, with a strict upper bound on the total
number of flavors (i.e., weakly coupled hypermultiplets). This comes about because in five dimensions,
supersymmetry constrains the metric on the Coulomb branch moduli space. To reach a conformal fixed point (starting from a gauge theory),
we need to be able to reach the singular regions of moduli space, but having more than
one gauge group factor obstructs this limit.

At first sight, this result
would seem to severely constrain the possible 5D SCFTs which can arise
from 6D SCFTs,
because the structure of many
stringy constructions appears to often take the form of a
quiver gauge theory, i.e., a gauge theory of precisely the type ruled out by
reference \cite{Intriligator:1997pq}. The key loophole \cite{Aharony:1997ju,Bergman:2012kr} is that
by moving in the vacuum moduli space of the 6D SCFT compactified on $S^1$,
one may reach points at which the effective 5D theory is superconformal.
While moving in the moduli space, one may reach a region in which
the inverse gauge coupling squared of the field theory is formally negative.
Before reaching such a region, the
effective field theory description which had been valid in the gauge theory
region breaks down  and undergoes a phase transition. While such an operation is ill-defined in gauge theory, it has a well-known
meaning in Calabi--Yau geometry: It is a flop transition! In M-theory
compactified on a Calabi--Yau threefold, flopping a curve formally means we continue its area
 to a negative value. What is really happening is that we pass
 from one chamber of K\"{a}hler moduli space to another and
 the curve being flopped is the one whose area controls the value of the inverse gauge coupling squared. In the
flopped phase we get another Calabi--Yau geometry. In the 5D SCFT literature this is
sometimes referred to as a ``UV duality,'' though we shall avoid this terminology.

In this paper we study the phase structure of 5D theories which descend from
compactification of a 6D SCFT or its deformations. For some preliminary analyses
of these theories, see e.g. \cite{DelZotto:2015rca,Ohmori:2015pia,Witten:1996qb}.
One of the general lessons from \cite{Heckman:2015bfa} is that an appropriate partial tensor
branch of a 6D SCFT is just a generalization of a quiver gauge theory in which the
link fields are themselves strongly coupled 6D\ SCFTs. Geometrically, the
tensor branch is obtained by performing a partial resolution of collapsing
curves in the base of the elliptic fibration. Starting from this partial
tensor branch, reduction on a circle takes us to a generalization of a 5D
quiver gauge theory. Alternatively, we can remain at the 6D fixed point
and reduce on a circle. For $(1,0)$ theories, we find that this always yields a 5D SCFT, or more precisely, a collection of between one and four 5D SCFTs.

\begin{figure}%
\centering
\includegraphics[
scale = 0.50, trim = 0mm 30mm 0mm 30mm]%
{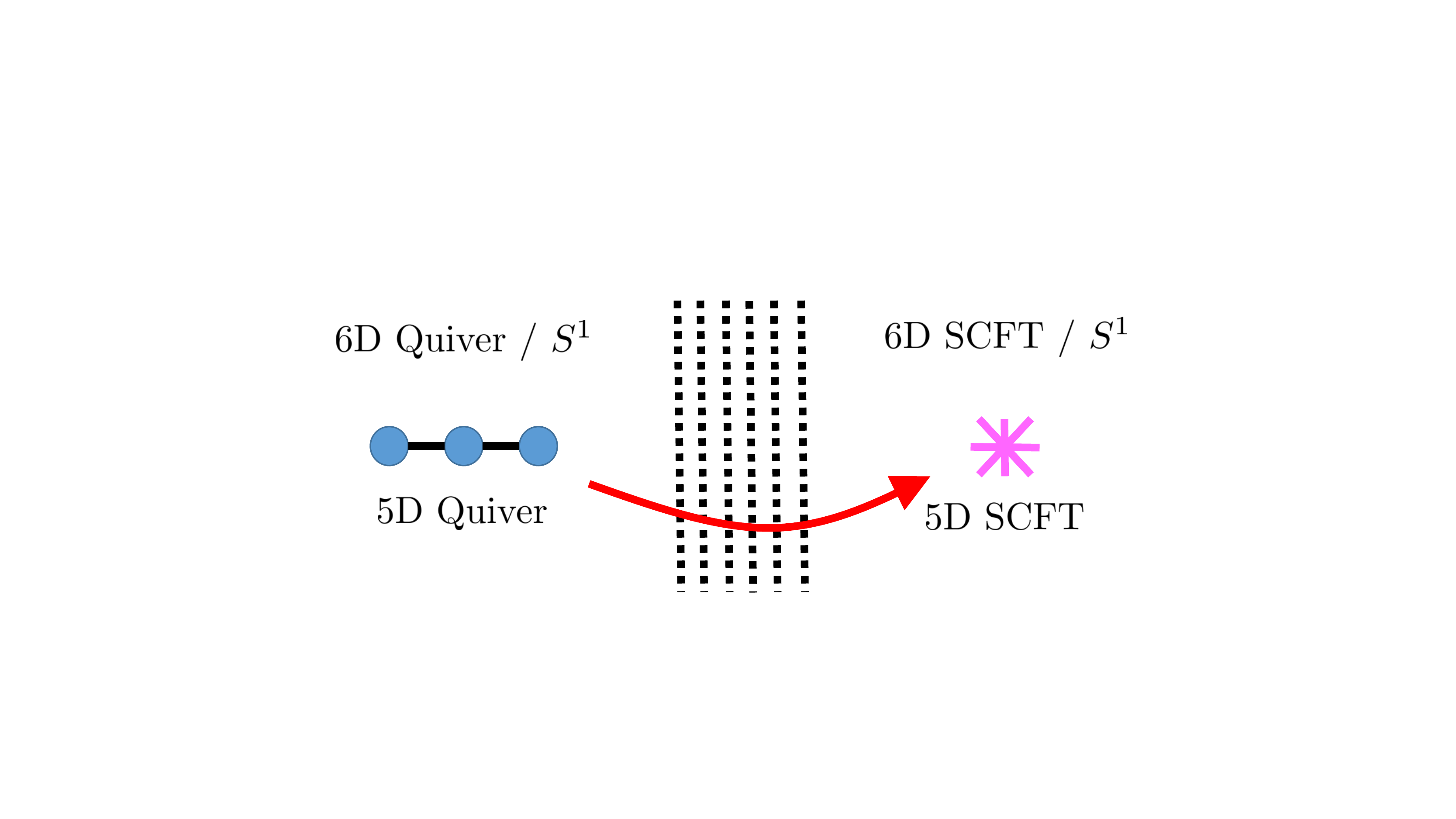}
\caption{Depiction of the phase structure for 6D theories reduced on a circle. Reducing a
$(1,0)$ 6D SCFT leads to a 5D SCFT, as indicated on the right. A sequence of flop transitions in the
extended K\"ahler cone of the Calabi--Yau threefold connects this chamber of moduli space to the one
obtained by dimensional reduction of the generalized 6D quiver. This leads to a generalized 5D quiver,
which need not possess a fixed point in this chamber of moduli space.}
\label{PhaseDiagram}%
\end{figure}

Our primary claim is that these two 5D theories are connected by a path in
moduli space which is in general realized by a sequence of flop transitions. To see this,
note that F-theory compactified on an elliptic Calabi--Yau threefold is, under
reduction on a further circle, described by M-theory on the same Calabi--Yau
threefold \cite{Vafa:1996xn,Morrison:1996na,Morrison:1996pp}.\footnote{In what follows we shall always assume a
Kaluza-Klein reduction on the circle in which we do not quotient by an automorphism of the
Calabi--Yau threefold. We also ignore potential ambiguities associated with the spectrum
of defects (see e.g. \cite{DelZotto:2015isa}).} In the M-theory description, the volume $V_{E}$ of the elliptic
fiber is related to the radius $R_{S^{1}}$ of the circle as:%
\begin{equation}
V_{E}=1/R_{S^{1}}.
\end{equation}
Compactification on a circle of the 6D tensor branch theory is realized by first resolving
the base of the F-theory model, and then resolving the elliptic fiber,
taking it to infinite size. Compactification of the 6D\ SCFT is realized by
only resolving the elliptic fiber taking it to infinite size. From the geometric engineering perspective, the latter possibility gives rise to a 5D SCFT because we automatically have divisors
collapsed to points. However, the geometry also indicates that the former is indeed a phase connected to the 5D SCFT.
We give a conceptual depiction of this trajectory in
Figure \ref{PhaseDiagram}.

So, whereas compactification of the 6D\ SCFT generates a 5D\ SCFT,
the generalized quiver will not necessarily lead directly to a 5D\ SCFT.
Rather, one must consider a motion in the extended K\"ahler
cone of the Calabi--Yau threefold. The existence of the F-theory model is what
guarantees that such a motion in moduli space is possible, and does indeed
lead to a non-trivial 5D fixed point.

We stress that the moduli space for M-theory on a CY three-fold used in a geometric engineering of a 6D SCFT within F-theory is strictly larger than the moduli space of a 5D SCFT: indeed, it equals the moduli space of the 6D SCFT compactified on $S^1$. To obtain the moduli space of the 5D SCFT, the radius of the circle must be taken to zero size. Correspondingly, $V_E$ must be taken to infinity. There are different inequivalent limits
in which the volume of the elliptic fiber is sent to infinity, leading to different 5D fixed points. This is somewhat reminiscent of what happens for 6D little string theories, that admit various
inequivalent decoupling limits, leading to distinct 6D SCFTs\cite{Bhardwaj:2015oru}.

From the perspective of M-theory compactified on a non-compact
Calabi--Yau threefold, generating a 5D\ SCFT simply requires that  some divisors  simultaneously
collapse  to a point at some location in the moduli space. There can
be multiple such locations, possibly located in distinct phase regions.

Of course, the above remarks prompt the question as to what fixed point is actually
realized by compactifying a 6D SCFT on a circle. Geometrically, we
characterize this singular limit by F-theory on a base $\mathbb{C}^{2}%
/\Gamma_{U(2)}$, with $\Gamma_{U(2)}$ a discrete group of $U(2)$. Only some
discrete subgroups lead to a consistent base for an F-theory model, and have
been classified in \cite{Heckman:2013pva} (see also \cite{Morrison:2016nrt}). Making such a choice, we construct a
Weierstrass model:
\begin{equation}
y^{2}=x^{3}+fx+g\text{,} \label{Weierstrass}%
\end{equation}
where here, $f$ and $g$ are polynomials in the holomorphic coordinates of
$\mathbb{C}^{2}$ which transform equivariantly under the action by the group
$\Gamma_{U(2)}$.
The order of vanishing for $f$ and $g$ dictates the enhancement for elliptic fibrations. This
characterization provides a direct way to access the 5D fixed point: Since we
have not performed any resolutions in the base, the only thing left for us to
do is take the limit where the elliptic fiber class expands to infinite size while remaining maximally singular.\footnote{ Naively, we can think of a given singular elliptic fiber as if it corresponds to an affine ADE graph $\widehat{\mathfrak{g}}$, the latter requirement amounts to taking the $V_E\to \infty $ limit sending  $\widehat{\mathfrak{g}} \to \mathfrak{g}$.} In this limit, we find that the 5D theory breaks up into at most four
decoupled SCFTs.  In particular,  the number of such constituent 5D SCFTs is much smaller than the
dimension of the tensor branch for the 6D SCFT. Some of these constituents
correspond to supersymmetric orbifold singularities of the form $\mathbb{C}%
^{3}/\Gamma_{SU(3)}$ for $\Gamma_{SU(3)}$ a finite subgroup of $SU(3)$. There
is typically another
constituent corresponding to collapsing a collection of four-cycles to a  non-orbifold singularity.

To illustrate these points, we also present a number of concrete examples.
Perhaps the simplest class of examples are those where the $\Gamma_{U(2)}%
$-equivariant polynomials $f$ and $g$ of equation (\ref{Weierstrass}) are
generic, i.e., no tuning is performed. These were referred to as
\textquotedblleft rigid theories\textquotedblright\ in reference \cite{Heckman:2013pva}.
For these theories, we can fully characterize the resulting 5D fixed point just
using the data of $\Gamma_{U(2)}$ itself. Further tuning leads us to
additional examples of generalized quivers, some of which admit a
rather simple form in F-theory. All of these cases lead to novel generalized
quiver gauge theories in five dimensions, and the F-theory model serves to
specify a path in moduli space to a fixed point after several flops.

The rest of this paper is organized as follows. First, in section \ref{sec:REVIEW} we give
a general review of how to generate 5D\ SCFTs from compactifications of
M-theory on a non-compact Calabi--Yau threefold. After this, we turn in section
\ref{sec:6D} to a brief review of the construction of 6D\ SCFTs via F-theory,
emphasizing the particular role of the orbifold singularity in the base. We
next turn in section \ref{sec:REDUCE} to an analysis of the 5D effective theories obtained
by directly compactifying a 6D\ SCFT on a circle, as well as the
compactification of its tensor branch deformation. We illustrate
these general points with specific examples in section \ref{sec:EXAMPLES},
and present our conclusions and some directions for future work
in section \ref{sec:CONC}. Additional details on the phases of the simple
rank one non-Higgsable clusters are presented in Appendix A.

As this paper neared completion, we received \cite{Hayashi:2017jze}
which considers a number of the same examples.  See also \cite{kim-private}.

\section{5D SCFTs from M-theory \label{sec:REVIEW}}

In preparation for our analysis of 6D theories compactified on a circle, in
this section we review the construction of 5D SCFTs via M-theory on a (non-compact)
Calabi--Yau threefold $X$.\footnote{ See e.g. \cite{Cadavid:1995bk,Ferrara:1996hh,Ferrara:1996wv}
for the case of a compact Calabi--Yau.} To realize an interacting
fixed point we need to reach a singular limit in Calabi--Yau moduli space, which
we expect to be resolved in the physical theory by the presence of additional massless / tensionless states.
Said differently, we expect 5D SCFTs for M-theory on any
 canonical singularity $P \in {X}_{\text{sing}}$ with a crepant resolution (i.e., Calabi--Yau blowup) $\pi \colon X \to {X}_{\text{sing}}$ which includes curve(s) and divisor(s)
in the inverse image $\pi^{-1}(P)$ \cite{Intriligator:1997pq}.

The geometric method we present subsumes other methods such as
the construction of 5D\ SCFTs via webs of $(p,q)$ five-branes in type
IIB\ string theory. Indeed, as is well-known, each of these web diagrams also
defines a toric Calabi--Yau threefold \cite{Leung:1997tw}.
The conformal limit in such constructions involves
bringing the various filaments of the web to the same location in the web,
i.e., a singular point, and in the interacting case always involves some
compact face of the $(p,q)$ web collapsing to zero size. In toric geometry,
such faces are interpreted as compact divisors, and the limit where the face
degenerates to zero size at a single point simply corresponds to the contraction of this divisor
to a point.

Let us now turn to the construction of M-theory on a canonical singularity and
explain in more general terms why we expect to realize 5D\ SCFTs. To see why,
recall that we measure volumes of
even-dimensional cycles by integrating powers of the K\"{a}hler form $J$. For
example, for a two-cycle $C$, the volume is:%
\begin{equation}
\text{Vol}(C)=\underset{C}{\int}J.
\end{equation}
For an M2-brane wrapped over a two-cycle, we get a BPS particle with mass proportional to this volume. For an M5-brane
wrapped over a divisor, we get a BPS string with tension specified by the volume of this divisor. In the limit
where the volume of the divisor passes to zero, this tension drops to zero.
A priori, the region in moduli space where
particles become massless and strings become tensionless can be
different \cite{Ganor:1996pc}.

Now, to generate an interacting fixed point, we
require at least one non-trivial divisor to collapse to a point in the geometry. The
reason is that with just collapsing curves, we only obtain some collection of
free hypermultiplets whereas with divisors collapsing to a curve, we get nonabelian gauge symmetry rather than an interacting fixed point. Assuming, then, that we have at least one collapsing
divisor, our task reduces to determining possible connected configurations of curves and divisors
which can all collapse simultaneously to a single point.

A necessary and sufficient condition for arranging this is to require first of all,
that we have a non-compact Calabi--Yau with a complete metric (i.e., we can decouple gravity), and second of all,
that the metric on the K\"{a}hler moduli space remains positive definite as we
pass to the putative singular point of moduli space.

For M-theory on a compact Calabi--Yau threefold $X$ with
$h^{1,1}$ K\"ahler moduli, if we choose a basis $D_I\in H^{1,1}_{\text{cpt}}(X)$, then the K\"ahler form is given by
\begin{equation}
J = \sum_{I=1}^{h^{1,1}} m^I D_I.
\end{equation}
Scaling the K\"ahler class does not change the M-theory
moduli, so the K\"ahler moduli are usually expressed as the ``volume one locus''
within $H^{1,1}(X)$, namely we use effective coordinates
\begin{equation}
\varphi^{I}\equiv m^{I}/V^{1/3}, \qquad I = 1, ..., h^{1,1}-1
\end{equation}
where $V \equiv \frac{1}{3!} \int_X J \wedge J \wedge J$. In practice we can scale $V$ to infinity and simultaneously rescale the $m^{I}$ in such a way that
\begin{equation}
\varphi^{I}=\underset{C_{I}}{\int}J, \qquad I = 1, ..., h^{1,1}-1
\end{equation}
remains finite and possibly non-zero. Here, $C_I$ is a the
basis of dual compact 2-cycles. An M2-brane wrapped over such a curve yields a BPS particle with
mass specified by $\varphi^I$. The bosonic superpartners of $\varphi$ define abelian vector bosons, which we
denote by $A^{I}$. They are given by integrating the three-form potential of
M-theory over the same two-cycles:%
\begin{equation}
A^{I}=\underset{C}{\int}\mathcal{C}_{(3)}.
\end{equation}
Similarly, one can introduce dual coordinates $\varphi_I \equiv D_{IJK} \varphi^J \varphi^K$ where $D_{IJK}$ is the triple intersection number of $X$, that controls the size of a basis of four-cycles of $X$. The $\varphi^{I}$ are the coordinates along the Coulomb phase which control the masses of BPS particles for the 5D theory, while the $\varphi_{I}$ are the dual coordinates, which control the tensions of the BPS monopole strings of the 5D theory.

The moduli space of M-theory on $X$ is given by the extended K\"ahler cone of $X$ \cite{Witten:1996qb}. A wall for a chamber of moduli space $\mathfrak{C}$ is defined by the condition that either (1) a curve shrinks to a point or a divisor shrinks to (2) a curve or (3) a point. For a given chamber $\mathfrak{C}$, the effective action for these abelian vector multiplets
is controlled by the 5D prepotential. Its form is given by a cubic polynomial in the
K\"ahler moduli:%
\begin{equation}
\mathcal{F}_{\mathfrak{C}}=\frac{1}{3!} D_{IJK} \varphi
^{I}\varphi^{J}\varphi^{K},
\end{equation}
where the $D_{IJK}$ are given by the triple intersection numbers for divisors
in the Calabi--Yau threefold:%
\begin{equation}
D_{IJK}=D_{I}\cdot D_{J}\cdot D_{K}.
\end{equation}
From this, we can read off the metric on moduli space:%
\begin{equation}
G_{IJ}=\frac{\partial^{2}\mathcal{F}_{\mathfrak{C}}}{\partial\varphi
^{I}\partial\varphi^{J}}.
\end{equation}
Indeed, the low energy effective action contains $h^{1,1}-1$ 5D abelian vector
multiplets
with couplings (see e.g. \cite{Intriligator:1997pq}):
\begin{equation}
L_{\text{eff}}= G_{IJ}d\varphi^{I}\wedge\ast d\varphi^{J}+G_{IJ}%
F^{I}\wedge\ast F^{J}+\frac{D_{IJK}}{24\pi^{2}}A^{I}\wedge F^{J}\wedge F^{K} + \cdots
\end{equation}
where here, $F^{I}=dA^{I}$ is the field strength for the vector boson.

Now, to reach a conformal fixed point, it is necessary for us to move to a
singular region of the geometry. So, we select some subset of the $\varphi
^{I}$, which we denote by the restricted index $\varphi^{i}$. We then hold
fixed the remaining K\"ahler moduli so that, for example, derivatives of the
prepotential with respect to these moduli are set to zero. $G_{ij}$ gives the matrix of effective gauge couplings, and with respect to
this subset, we demand that the $G_{ij}$ is positive away from the origin.
When this condition is satisfied, we can collapse the associated four-cycles
to zero size, and we thus expect to realize a 5D SCFT. When this condition is
not satisfied, we cannot simultaneously contract the size of all of the
divisors. From this perspective, the task of determining candidate SCFTs from
M-theory configurations involves analyzing all possible choices of divisors
subject to these criteria. This condition of positivity as we move to the
origin of moduli space can also be stated as a convexity condition on our
prepotential \cite{Intriligator:1997pq}:%
\begin{equation}
\mathcal{F}_{\mathfrak{C}}\mathcal{(\lambda}_{(1)}\varphi_{(1)}^{i}%
+\mathcal{\lambda}_{(2)}\varphi_{(2)}^{i})\leq\mathcal{F}_{\mathfrak{C}%
}\mathcal{(\lambda}_{(1)}\varphi_{(1)}^{i})+\mathcal{F}_{\mathfrak{C}}(\lambda_{(2)}%
\varphi_{(2)}^{i}) \label{convexity}%
\end{equation}
with:%
\begin{equation}
\mathcal{\lambda}_{(1)}+\mathcal{\lambda}_{(2)}=1\text{ \ \ and \ \ }%
0\leq\mathcal{\lambda}_{(1)}\text{, }\mathcal{\lambda}_{(2)}\leq1.
\end{equation}
If we cannot satisfy this criterion, then we conclude that it is not possible
to reach a conformal fixed point in a particular chamber.

In such situations, we can of course, also contemplate formally continuing
some of the parameters $\varphi^{I}$ to negative values, i.e., we allow
negative area for a given curve. Geometrically this is described by a flop
transition between two Calabi--Yau manifolds with the same Hodge numbers. In
this flopped phase, the structure of the triple intersection numbers will
change, and consequently, also the prepotential. Observe that an M2-brane wrapped on such a curve will generate a
BPS state with mass which goes from being positive
to negative.\footnote{Many flop transitions can be thought of as being realized by replacing a given curve with normal bundle either $\mathcal{O}(-1)\oplus \mathcal{O}(-1)$ or $\mathcal{O}\oplus \mathcal{O}(-2)$ with an $\mathbb{F}_1$ which is then shrunk down with respect to the other ruling \cite{friedman1983birational}.  However, there are also flops on rational curves whose normal bundle is $\mathcal{O}(1)\oplus \mathcal{O}(-3)$
\cite{[L],[P],pagoda,gorenstein-weyl,curto-morrison}.}
Once we have the new triple intersection numbers, we can again analyze whether
the prepotential is convex in the new chamber $\mathfrak{C}_{\text{new}}$. An
important feature of the new prepotential is that it retains much of the
structure of the original. To exhibit this, we view
$\mathcal{F}_{\mathfrak{C}}$ as a function of positive values for the moduli
$\left\vert \varphi^{i}\right\vert $. The change between the prepotential for the
old and new phase can be written in the form
\begin{equation}
\mathcal{F}_{\text{new}}-\mathcal{F}_{\text{old}}=\frac1{3!}\mathcal{L}^3,
\end{equation}
where $\mathcal{L}=(\sum m^ID_I )\cdot C_{\text{flop}}$ is a linear function vanishing on the wall between
the two K\"ahler cones which is positive after the flop\cite{beyond,Morrison:1996xf}.

An interesting open question is to provide an explicit classification of all
canonical singularities which can generate 5D\ SCFTs. Compared with the
classification strategy for 6D SCFTs generated by F-theory \cite{Heckman:2013pva, Heckman:2015bfa},
this is a far more intricate question because it involves tracking the collapse of
four-cycles in our geometry. For example, we generate canonical singualarities
from the orbifolds $\mathbb{C}^{3}/\Gamma_{SU(3)}$ with $\Gamma_{SU(3)}$ a finite
subgroup of $SU(3)$. The resolved geometry will typically contain multiple
divisors all collapsing to zero size simultaneously. There can also be various
intermediate limits where a K\"{a}hler surface first collapses to a curve, and
then this curve futher degenerates to a point. In some cases, this
degeneration has an interpretation in terms of 5D gauge theory, though in most
cases it is more \textquotedblleft exotic\textquotedblright\ from the
perspective of effective field theory.

Our plan in the rest of this section will be to illustrate some of these
considerations for a few well known examples. We will then proceed in the
following sections to a much broader class of examples as engineered by
compactifications of 6D\ SCFTs on a circle.

\subsection{Single Divisor Theories}

In this subsection we consider 5D SCFTs generated by a single collapsing
divisor in a Calabi--Yau threefold. Assuming that the normal geometry in the Calabi--Yau
threefold is smooth, we can locally characterize the geometry by the total
space $\mathcal{O}(K_{S})\rightarrow S$, with $S$ the K\"{a}hler surface. The
triple intersection number for the divisor $S$ can also be evaluated using
intersection theory on the surface itself.\ Indeed, we have:%
\begin{equation}
S\cdot_{\text{CY}}S\cdot_{\text{CY}}S=K_{S}\cdot_{S}K_{S},
\end{equation}
where the subscripts for $\cdot_{\text{CY}}$ and $\cdot_{S}$ indicate that the
intersection takes place in the corresponding K\"{a}hler manifold. A necessary
condition to reach a conformal fixed point is that the metric on the moduli
space remains positive definite, so we must require:%
\begin{equation}
K_{S}\cdot K_{S}>0\text{.} \label{Kpos}%
\end{equation}
This condition is somewhat milder than the condition that we can directly
contract $S$ to a point. Indeed, to decouple gravity in a local M-theory
model, we either require $S$ to contract to a point, or to a curve. In the
former case, we impose the stronger condition $-K_{S}>0$, which restricts
us to the del Pezzo surfaces. A milder condition is that $K_{S}\cdot K_{S}>0$.
This is satisfied, for example, for the Hirzebruch surfaces $\mathbb{F}_{n}$,
with $n\geq2$ (which are not Fano). Observe that condition (\ref{Kpos}) is not
satisfied for a del Pezzo 9 (i.e., half K3) or K3 surface.

Now, in the case of the del Pezzo $k$ surfaces $dP_{k}$, i.e., $\mathbb{P}^{2}$
blown up at a $0\leq k\leq8$ points, there is a well known correspondence for
$k\geq1$ to a 5D $SU(2)$ gauge theory with $k - 1$ hypermultiplets. In this
geometric picture, the $SU(2)$ gauge theory is realized by noting that each
del Pezzo surface can also be viewed as a $\mathbb{P}_{\text{fiber}}^{1}$
bundle over a $\mathbb{P}_{\text{base}}^{1}$, possibly with some locations
where this fibration degenerates. In the limit where the fiber $\mathbb{P}%
_{\text{fiber}}^{1}$ collapses to zero size, we get a curve of $A_{1}$
singularities, realizing an $SU(2)$ gauge theory. The locations where the
fibration degenerates lead to local enhancements in the singularity type,
providing additional matter fields \cite{Morrison:1996xf,Douglas:1996xp}. The case $k=0$ does
not admit an interpretation as an $SU(2)$ gauge theory, but is instead known
as the \textquotedblleft$E_{0}$ theory,\textquotedblright\ (or $\mathbb{C}^3/\mathbb{Z}_3$) as in reference
\cite{Morrison:1996xf}. In all cases, we reach a conformal fixed point by
collapsing the K\"ahler surface to a point. This also leads to an enhancement in
the flavor symmetry, which can be directly computed via the geometry \cite{Morrison:1996xf}.
It is given by the exceptional group $E_{k}$, where for $k<6$ we simply delete
appropriate nodes from the affine Dynkin diagram $\widehat{E}_{8}$.

A more unified perspective on all of these examples comes from first starting
with the local geometry defined by a del Pezzo nine surface \cite{Morrison:1996pp,Morrison:1996xf}.
This can be viewed as $\mathbb{P}^{2}$ blown up at nine points, and is also described by a
Weierstrass model of the form:%
\begin{equation}
y^{2}=x^{3}+f_{4}x+g_{6},
\end{equation}
namely, we have an elliptic fibration over a $\mathbb{P}^{1}$ in which the
Weierstrass coefficients $f_{4}$ and $g_{6}$ are respectively degree four and
six homogeneous polynomials. Flopping the zero section of this model, we then
blow down additional points to reach the various del Pezzo models. These
correspond in the field theory to adding mass deformations to the associated hypermultiplets.

An additional class of examples are given by the Hirzebruch surfaces
$\mathbb{F}_{n}$, which for $n>1$ are not Fano, i.e., $-K_{S}$ is not
positive. From the perspective of the M-theory construction, we cannot
construct a local metric which is complete. From a field theory point of view,
this is the statement that there is no way to fully decouple gravity. Rather,
we must include some additional degrees of freedom to complete the
description. In the geometry, this requires us to introduce some additional
divisors. Assuming the existence of at least one more divisor, we can now see
why such a model could produce a 5D\ SCFT. First of all, we recall that
$\mathbb{F}_{n}$ can also be viewed as a $\mathbb{P}_{\text{fiber}}^{1}$
bundle over a $\mathbb{P}_{\text{base}}^{1}$, in which the first Chern class
of the bundle is $n$. If we can take a limit in the Calabi--Yau moduli space in which
the volume of $\mathbb{P}_{\text{base}}^{1}$ collapses to zero size, we get a weighted
projective space $\mathbb{P}_{[1,1,n]}^{2}$. This can then collapse to zero
size. Of course, this assumes that we can collapse the $\mathbb{P}%
_{\text{base}}^{1}$ to zero size, and this in turn assumes that this curve is
a subspace of another K\"{a}hler surface in the geometry. The condition we are
thus finding is that this other surface must also collapse to zero size.

\subsection{Quiver Gauge Theories}

So far, we have focussed on the geometric construction of 5D\ SCFTs. One can
also attempt to engineer examples using methods from low energy effective
field theory. Along these lines, we can consider a 5D\ quiver gauge theory
with simple gauge group factors $G_{1},...,G_{l}$, and with matter fields in
some representation between these gauge group factors, i.e., hypermultiplets in
bifundamental representations $(R_{i},R_{j})$. The construction of such models
is concisely summarized by a quiver diagram.

Geometrically, we engineer a 5D gauge theory with gauge group $G$ by
introducing a curve of singularities. Locally, these are described by
specifying a curve, and then taking a fibration by a space $\mathbb{C}%
^{2}/\Gamma_{\text{ADE}}$ with $\Gamma_{\text{ADE}}$ a discrete subgroup of
$SU(2)$ \cite{Katz:1997eq}. This yields the ADE\ groups, and the non-simply laced algebras can
also be realized by allowing suitable monodromies in the fibration \cite{Bershadsky:1996nh}. In these
models, the value of the gauge coupling is controlled by the volume of the
base curve. We can also engineer matter fields by introducing local
enhancements in the singularity type of the fibration \cite{Katz:1996xe}.

Collisions between curves supporting gauge groups can also produce a strongly
coupled version of a hypermultiplet which is the 5D version of 6D
conformal matter \cite{DelZotto:2014fia}. Some canonical examples of such
behavior include the reduction of 6D conformal matter on a circle, a point we return
to shortly. In five dimensions one can also contemplate more intricate intersection
patterns, leading to further generalizations for 5D conformal matter.

Using methods either from gauge theory and/or geometry, it is possible to
calculate the prepotential for these sorts of models. A perhaps surprising
feature of all of these cases is that only for a single simple gauge group
factor do we have a chance of realizing a 5D\ SCFT connected to every chamber of moduli space\cite{Intriligator:1997pq}.
The reason for this is clear from the structure of the prepotential $\mathcal{F}$,
which contains a term of the schematic form:%
\begin{equation}
-\frac{1}{12}\left\vert c\varphi+\varphi^{\prime}\right\vert ^{3},
\end{equation}
where $\varphi$ is the Coulomb branch parameter(s) associated with one simple
gauge group factor, and $\varphi^{\prime}$ are associated with other Coulomb
branch parameters. Physically, the vevs of $\varphi^{\prime}$ can be viewed as
giving masses to some of the hypermultiplets. The issue is that the
contribution from such a term violates the convexity condition of line
(\ref{convexity}). Indeed, in the geometry, what is happening is that a curve
$C$ in a surface $S$ is collapsing to zero volume before that surface can
pass to zero volume as well. To continue the contraction of the surface, it is
thus necessary to assume that we can continue the volume of $C$ to
formally negative values, i.e., we must require the existence of a flop transition, bringing us to a different chamber of moduli
space.\footnote{As an example of this type, ref.~\cite{Aharony:1997ju} considers a $(p,q)$-fivebrane web construction of $SU(2)\times SU(2)$ gauge theory with a hypermultiplet
in the bifundamental representation.  In the associated Calabi--Yau geometry,
the flopped phase corresponds to $SU(3)$ gauge theory with two flavors in
the fundamental representation.  In general, however, one should not expect
the flopped phase of a gauge theory to again be a gauge theory.}

Without further input, we cannot conclude whether it is
possible to reach a 5D SCFT through a sequence of flops. What we can conclude,
however, is that in the chamber of moduli space where a quiver
gauge theory description is valid, we do not expect to reach a 5D\ SCFT. One
of our aims in this paper will be to elaborate on when we expect to achieve a
sequence of flop transitions to a chamber which supports a 5D\ SCFT.

In the case of 6D theories on $S^1$ the existence of such chamber is guaranteed from the existence of the 6D fixed point. To gain further insight into the structure of possible 5D\ SCFTs, we shall use this higher-dimensional perspective. This will help us in determining candidate 5D theories, as well as establishing the existence of flops between these models.

\subsection{M-theory on an Elliptic Calabi--Yau Threefold}

When approaching the construction of 5D SCFTs from a 6D origin, we must
consider M-theory on an elliptic Calabi--Yau threefold.  The threefold
need not be compact, but it should contain compact elliptic curves.

Specifically, we consider a proper\footnote{This means that the inverse
image of any point is compact.}  map $\pi:X\to B$ from a (non-compact)
Calabi--Yau
threefold to a (non-compact) surface $B$ whose general fiber
is a compact elliptic curve.  We assume that there is a birational
section\footnote{For our present purposes, a birational multi-section
would work equally well, at the expense of a more complcated notation.}
of this fibration $\sigma: \widetilde{B} \to X$, where $\widetilde{B}\to B$
is an appropriate blowup.  Typically, we will consider bases $B$ which
are neighborhoods of a connected collection of compact curves, but our analysis
will also hold more generally.

We are interested in the K\"ahler parameters of $X$.  This is not really
a well-defined question, because when $X$ is non-compact one can imagine
different boundary conditions for the metric.  However, there are
certain K\"ahler parameters which are visible in our setup, and they
are measured by the areas of all of the compact curves on $X$.

More explicitly, we consider $Ch_1(X)$, the ``Chow group'' of
algebraic $1$-cycles, i.e.,
$\mathbb{Z}$-linear
combinations of irreducible compact
curves, modulo algebraic equivalence.  The equivalence relation is
generated by families of compact curves parameterized by a (possibly
non-compact) curve, in which singular fibers in the family are represented
by the corresponding linear combination of components weighted by
multiplicity.

The vector space of possible areas of elements of the Chow group
provides a description of
the space spanned by K\"ahler classes on $X$ having some fixed type of
boundary conditions.
We expect that for the families we study, after performing an appropriate
scaling on the base $B$ there are complete metrics on both $B$ and
$X$ with appropriate growth conditions at infinity which would nail down
the K\"ahler classes more precisely.

The K\"ahler classes themselves will be elements of the dual vector space
of $Ch_1(X)$,
or more precisely, of a cone within the dual vector space consisting of
all classes such that the area of any effective $1$-cycle is positive.
Compact divisors on $X$ will naturally give rise to elements of
the dual vector space, but in general, we may need non-compact divisors
as well as compact ones in order to fully describe the cone of
K\"ahler classes.

As in the case of compact $X$, the boundaries of the K\"ahler cone indicate
places where one or more curve classes shrink to zero area.  One way this can
come about is if the entire space $X$  shrinks to zero volume (by shrinking
 the fibers of an elliptic fibration or of a fibration by surfaces with trivial canonical
bundle, or by shrinking all of $X$ to a point.)  The only other
way this can come about is if a compact cycle on $X$ shrinks to a cycle
of lower dimension.

In the case of a finite collection of curves shrinking to points, it is sometimes
possible to find a ``flop'' which allows the K\"ahler moduli to be
continued past the boundary.  In this case, the flopped Calabi--Yau
has a K\"ahler cone of its own which meets the orignal cone along
a common part of the boundary.  Including all such cones gives the
``extended K\"ahler cone'' of $X$.

We will assume that $B$ is either a neighborhood of a singular point,
or else a neighborhood of a contractible collection of curves.  In this
case, we can expect gravity to decouple after an appropriate scaling limit.

To study possible emergent 5D SCFTs from this geometry, we wish to pass to
a limit in which the area of the elliptic curve goes to infinity.
(For fibers of $\pi$ which have more than one component, at least one
of those components must also go to infinite area, and more than one
may do so.)
By varying the K\"ahler cone and/or varying the choice of which components
of fibers go to infinite area,
there can be distinct limiting 5D theories, each obtained
by integrating out the very massive particles arising from an M2-brane
wrapped on the elliptic curve (or chosen components of fibers), when the area
is extremely large.  These distinct limiting theories
cannot be connected to each other directly in 5D
without re-introducing an elliptic
fiber.  We will see explicit examples of this phenomenon later in the
paper.

\section{F-theory on a Circle \label{sec:6D}}

To facilitate our understanding of 5D theories, and their possible conformal
fixed points, our aim in this section will be to turn to a higher-dimensional
perspective as provided by 6D\ SCFTs. The main tool at our disposal is the
recent classification of 6D\ SCFTs via F-theory compactification.
Along these lines, we shall first present some of the
salient features of these classification results.

We generate 6D\ SCFTs by working with elliptically
fibered Calabi--Yau threefolds over a non-compact base $B$. This is specified
by a Weierstrass model of the form:%
\begin{equation}
y^{2}=x^{3}+fx+g
\end{equation}
where $f$ and $g$ are sections of $\mathcal{O}(-4K_{B})$ and $\mathcal{O}%
(-6K_{B})$, respectively. Assuming we have such a Calabi--Yau threefold, the
condition to reach a 6D\ SCFT is that some subset of curves in the base can
simultaneously contract to zero size. This requires the intersection pairing
for these curves to be a negative definite matrix.\ Classification of
6D\ SCFTs thus proceeds in two steps. First, we seek out all possible
candidate bases $B$ which can support a 6D\ SCFT, and second, we classify all
possible elliptic fibrations over a given choice of base. The conformal fixed
point corresponds to the limit in which we collapse all curves to zero size.

Now, an important feature of this classification scheme is that the structure
of the bases take a quite restricted form in the limit where all curves
collapse to zero size, namely, the base is an orbifold singularity of the form
$\mathbb{C}^{2}/\Gamma_{U(2)}$ for $\Gamma_{U(2)}$ a discrete subgroup of
$U(2)$. An additional intriguing feature which is still only poorly understand
is that only specific finite subgroups of $U(2)$ are actually compatible with
the condition that we have an elliptically fibered Calabi--Yau threefold.

The geometry of 6D\ SCFTs can thus be understood in complementary ways. On the
one hand, we can consider the resolved phase where all curves are of finite
size, with volumes $t^{I}>0$ for the different two-cycles. This is referred to
as the tensor branch of the theory. On the other hand, we can pass back to the
conformal fixed point by collapsing all of these curves to zero size, i.e., we
take the limit $t^{I}\rightarrow0$.

Now, our interest in this paper will be on the types of 5D theories obtained
by compactifying our 6D theories on a circle of radius $R_{S^1}$.
The 5D BPS mass of a string wrapped on the $S^1$ is given by
$R_{S^1} \times t^I$.
Once we compactify on a circle, we reach M-theory on the same Calabi--Yau
threefold, but now the volume of the elliptic fiber is a physical
parameter, and identified with the inverse radius of the circle
compactification:%
\begin{equation}
V_{E}=1/R_{S^{1}}.
\end{equation}
Our expression for the 5D BPS mass can then be written as
$t^I/V_E$.  The decoupling limit needed to reach a 5D SCFT always
requires $V_E\to\infty$, but clearly this limit depends on
the behavior of these ratios.
Different choices of the ratios
correspond to different regions in the extended K\"ahler
cone of the Calabi--Yau threefold.
One choice is to take all $t^I=0$, which we view as the direct reduction
of the 6D SCFT.  Another choice corresponds to keeping some of the ratios
$t^I/V_E$ finite which is the reduction of a partial tensor branch from 6D.
These are of course connected by flop
transitions, but a priori, they could have very different chamber structures,
and may possess different degenerations limits which can support a 5D\ SCFT.

Let us consider the structure of each of these branches, as well as their
dimensional reduction on a circle. On the tensor branch of the 6D theory, we
have at least as many independent 6D tensor multiplets as simple gauge group
factors. In fact, one of the lessons from the classification results of
reference \cite{Heckman:2013pva, DelZotto:2014hpa, Heckman:2015bfa}
is that typically, many such extra tensor multiplets should be viewed as defining a generalization of hypermultiplets
known as \textquotedblleft conformal matter.\textquotedblright\ For example, a
configuration of curves in the base intersecting as:%
\begin{equation}
\lbrack E_{8}]1,2,2,3,1,5,1,3,2,2,1[E_{8}] \label{E8confmatt}%
\end{equation}
consists of eleven tensor multiplets, one associated with each curve. Here,
the notation $m, n$ refers to a pair of curves of self-intersection $-m$ and
$-n$ intersecting at one point. The entries in square brackets at the left and
right denote flavor symmetries for the 6D system. For each such curve, there
is minimal singularity type in the elliptic fibration over each curve, as
dictated by the structure of non-Higgsable clusters \cite{Heckman:2013pva, Morrison:2012np}.

The dimensional reduction of this system will consist of a number of 5D gauge
group factors, associated with their 6D counterparts, as well as additional
$U(1)$ gauge group factors coming from the reduction of the 6D tensor
multiplet to five dimensions. There is also rich collection of 5D Chern-Simons
terms coming from reduction of the associated 6D\ Green-Schwarz terms, and one
loop corrections (see e.g. \cite{Intriligator:1997pq, Grimm:2015zea}).

Instead of resolving all of the curves to finite size, we can also consider
mixed branches where only some of the curves are of finite size. This leads to
the notion of a generalized quiver gauge theory, with, for example,
exceptional gauge groups and conformal matter suspended between these gauge
group factors. For example, in line (\ref{E8confmatt}) we can
collapse all eleven intermediate curves to zero size, producing $E_{8}\times
E_{8}$ conformal matter.\ We can also gauge these flavor symmetries, i.e.,
place these factors on compact curves, and continue adding additional
conformal matter factors. Such generalized
quivers consist of a single linear chain of such D- and E-type gauge group
factors, with the rest interpreted as conformal matter. The conformal matter
sector can also be visualize as M5-branes probing an ADE singularity \cite{DelZotto:2014hpa, Heckman:2014qba}.

The dimensional reduction of such conformal matter sectors leads to well-known
5D gauge theories. For example, for an M5-brane probing an ADE\ singularity,
we obtain, at low energies, a D4-brane probing an ADE singularity, i.e., we
obtain an affine quiver gauge theory with gauge groups given by the Dynkin
indices of the gauge group factors. This system possesses a $G_{L}\times
G_{R}$ flavor symmetry (see e.g.\ \cite{Tachikawa:2015mha,Zafrir:2015uaa,Yonekura:2015ksa}), so  we can after
passing through an appropriate flop transition to reach a 5D CFT, also
view  this  as a type of 5D conformal matter for
the weakly gauged sector. Since 6D SCFTs have the form of generalized quivers,
we see that the reduction of the partial tensor branch leads to a similar
generalization of quiver gauge theories in 5D as well.
See section \ref{sec531} for further discussion.

Finally, we come to the last possibility where we do not resolve any of the
curves in the base of the fibration, and compactify the 6D\ SCFT\ directly on
a circle. In this case, we always expect to generate a 5D SCFT, since we have
divisors already collapsed to zero size.\footnote{The caveat to this
statement, is of course, the 6D $(2,0)$ theories because in this case the
geometry is of the form $\mathbb{C}^{2}/\Gamma_{SU(2)}\times T^{2}$, so there
are no collapsing divisors in the non-compact Calabi--Yau threefold.}

\section{6D\ SCFTs on a Circle \label{sec:REDUCE}}

In this section we study in detail the region of moduli space which in
most cases leads to a 5D fixed point, i.e., the dimensional reduction of a
$(1,0)$ 6D\ SCFT on a circle. In this case, we always aim to decompactify the
elliptic fiber first, leaving all other curves collapsed at zero size. In addition
to curves on the base, this would include all but one component of any (singular)
elliptic fiber.  Since the
base of our F-theory SCFT is already described by a collection of contractible
curves in the base, the presence of a collapsing $\mathbb{P}^{1}$ (as one
of the components corresponding to
a singular elliptic fiber) automatically generates a collapsing divisor
and thus a 5D fixed point in the associated M-theory compactification.\footnote{Here we
do not consider possible twists along the circle by the automorphisms of the Calabi-Yau.}

To characterize these 5D\ fixed points, it will prove convenient to adopt a
somewhat different perspective on the structure of our 6D\ SCFTs.\ Rather than
working with a quiver description corresponding to a base in which we have resolved all curves to finite size, we can
instead treat the base $B$ as an orbifold $\mathbb{C}^{2}/\Gamma_{U(2)}$, and
with the coordinates $x,y,f$ and $g$ of the Weierstrass model treated as
appropriate $\Gamma_{U(2)}$-equivariant sections of bundles on this
orbifold \cite{DelZotto:2014fia,Bertolini:2015bwa,Morrison:2016nrt}.
We specify the group action by the defining two-dimensional representation on
the holomorphic coordinates $s$ and $t$ of the covering space $\mathbb{C}^{2}%
$.\ (We consider only group actions on $\mathbb{C}^2$ in which the only fixed point for any
non-identify element of the group is the origin.)  To specify a Weierstrass model over this base, we choose to work in a
twisted\footnote{Similar considerations would also apply if we had instead
presented the Weierstrass model in a weighted projective space.}
$\mathbb{P}^{2}$ with homogeneous coordinates $[x,y,z]$ so that we have the
presentation:%
\begin{equation}
y^{2}z=x^{3}+f(s,t)xz^{2}+g(s,t)z^{3},
\end{equation}
where $f(s,t)$ and $g(s,t)$ are polynomials in the holomorphic coordinates $s$
and $t$ of the covering space $\mathbb{C}^{2}$. It is a twisted $\mathbb{P}%
^{2}$ in the sense that $[x,y,z]$ transform non-trivially under the group
action, and $f$ and $g$ transforming as sections of $\mathcal{O}(4K_{B})$ and
$\mathcal{O}(6K_{B})$. For $\gamma\in\Gamma_{U(2)}$, the transformation rules
are:
\begin{align}
\lbrack x,y,z]  &  \mapsto[\det(\gamma)^{2}x,\det(\gamma)^{3}%
y,z]\label{paction}\\
f(s,t)  &  \mapsto\det(\gamma)^{4}f(s,t)\label{faction}\\
g(s,t)  &  \mapsto\det(\gamma)^{6}f(s,t). \label{gaction}%
\end{align}
We wish to emphasize that it is necessary to take the orbifold of the
twisted $\mathbb{P}^2$ (and the Weierstrass hypersurface within it)
by the finite group $\Gamma_{U(2)}$.

In order to study this orbifold, we should consider the three standard
coordinate charts of the twisted $\mathbb{P}^2$.  One of these is the
\textquotedblleft standard\textquotedblright\ one for analysis of the
Weierstrass model, i.e., $z=1$, and the others are at $x=1$ and  at $y=1$:%
\begin{align}
y^{2}  &  =x^{3}+f(s,t)x+g(s,t) && z=1\text{ patch}\label{zonepatch}\\
y^2z  &  =1+f(s,t)z^{2}+g(s,t)z^{3} && x=1\text{ patch.}\label{xonepatch}\\
z  &  =x^{3}+f(s,t)xz^{2}+g(s,t)z^{3} && y=1\text{ patch.}
\label{yonepatch}%
\end{align}
The first remark is that in the $x=1$ patch, it is not possible for $z$
to vanish at any point on the hypersurface.  Thus, all the points on
the hypersurface in the $x=1$ patch also lie in the $z=1$ patch and
we  need not  consider the $x=1$ patch any further.

Consider next the
$y=1$ patch. Here, we see that the
hypersurface is smooth near $z=0$, due to the linear term in $z$ on the lefthand side of
the defining hypersurface equation. On this chart, the group action on the
affine coordinates is:%
\begin{equation}
(s,t,x,z)\mapsto(\gamma_{11}s+\gamma_{12}t,\gamma_{21}s+\gamma_{22}%
t,\det(\gamma)^{-1}x,\det(\gamma)^{-3}z),
\end{equation}
where in the first two entries, we have indicated the entries of the group
element $\gamma$ in the defining representation. Since we are solving for $z$
in line (\ref{yonepatch}), the action on $z$ is the same as that on the equation, and the geometry is locally characterized (near $z=0$) as having a quotient
singularity of the form $\mathbb{C}^{3}_{s,t,x}/\Gamma_{SU(3)}$ where the explicit
group action decomposes into a block structure of the form:%
\begin{equation}
\gamma_{SU(3)}=\left[
\begin{array}
[c]{cc}%
\gamma_{U(2)} & \\
& \det(\gamma_{U(2)})^{-1}%
\end{array}
\right]  ,
\end{equation}
in the obvious notation. This  gives a 5D SCFT when $\Gamma_{U(2)}$ is non-trivial.

From this, we already see an interesting prediction
from the geometry: when the determinant map%
\begin{equation}
\det:\Gamma_{U(2)}\rightarrow U(1),
\end{equation}
has a non-trivial kernel, the singularity is not isolated, and we also expect
a non-trivial flavor symmetry.  The flavor symmetry is the algebra of type A, D, or E corresponding to the kernel of $\det$, which is a subgroup of
$SU(2)$.
In principle, of course, this may only be a subalgebra of the full
flavor symmetry of the 5D theory.

Turning now to the $z=1$ patch,
we need to analyze fixed points of the orbifold action.  In this patch,
the action on affine coordinates is
\begin{equation}
(s,t,x,y)\mapsto(\gamma_{11}s+\gamma_{12}t,\gamma_{21}s+\gamma_{22}%
t,\det(\gamma)^{2}x,\det(\gamma)^{3}y),
\end{equation}
where again in the first two entries, we have indicated the entries of the group
element $\gamma$ in the defining representation.
The origin is a codimension four fixed point for the group action on the
affine coordinates, so if the origin lies on the
hypersurface it provides one of the singular points.

The codimension three locus $s=t=y=0$ is fixed by the kernel of $\det^2$,
the codimension three locus
$s=t=x=0$ is fixed by the kernel of $\det^3$, and the codimension two locus
$s=t=0$ is fixed by the kernel of $\det$.  To determine which of
these loci intersect the hypersurface away from the origin,
we examine the Weierstrass equation.  We have already discussed this
in the case of the kernel of $\det$, which leads to a fixed curve
within the hypersurface and a flavor symmetry whose type is determined
by the subgroup $\ker(\det)\subset SU(2)$.

In order for $s=t=x=0$ to intersect the hypersurface away from the
origin, we must have $g(0,0)\ne0$.
In order for $s=t=y=0$ to intersect the hypersurface away from the
origin, we must have either $f(0,0)\ne0$ or $g(0,0)\ne0$.  And finally,
in order for $s=t=0$ to intersect the hypersurface away from the
origin, we must have either $f(0,0)\ne0$ or $g(0,0)\ne0$.
Thus, whenever there is a fixed point away from the origin
we may assume that $\det^4=1$ or $\det^6=1$.  Let us consider the possibilities
one at a time.

First, if $\det=1$ then
the only singularity away from the origin is the non-isolated one.

Next, if $\det^2=1$ and the polynomials are generic, then $f(0,0)\ne0$
and $g(0,0)\ne0$.  The action of $\Gamma_{U(2)}$ on the elliptic curve
is multiplication by $-1$, with three fixed points at the zeros of
$x^3+f(0,0)x+g(0,0)$ (with $y=0$) and a fourth at infinity.

If $\det^3=1$ and the polynomials are generic, then $g(0,0)\ne0$ but
$f(0,0)=0$.  The action of $\Gamma_{U(2)}$ on the elliptic curve is
by an automorphism of order three, which has two fixed points at
$(x,y)=(0,\pm\sqrt{g(0,0)})$ and a third at infinity.

If $\det^4=1$ and the polynomials are generic, then $f(0,0)\ne0$ but
$g(0,0)=0$.  The action of $\Gamma_{U(2)}$ on the elliptic curve is
by an automorphism of order four; on the quotient, we have
the fixed point $(x,y)=(0,0)$
with stabilizer $\Gamma_{U(2)}$ and one fixed point with stabilizer
$\ker (\det^2)$
(coming from the two points $(x,y)=(\pm\sqrt{-f(0,0)}$ which are exchanged
by the action),
as well as the point at infinity.

Finally, if $\det^6=1$ and the polynomials are generic, then
$g(0,0)\ne0$ but
$f(0,0)=0$. The action of $\Gamma_{U(2)}$ on the elliptic curve is
by an automorphism of order six. On the quotient,
the origin is a fixed point with
stabilizer $\Gamma_{U(2)}$;
there is one fixed point with stabilizer
$\ker (\det^3)$ (coming from the two points $(x,y)=(0,\pm\sqrt{g(0,0)})$
which are exchanged by the action), and
one with stabilizer $\ker (\det^2)$ (coming from the three points
$(x,y)=(e^{2\pi ik/3}\, \sqrt[3]{-g(0,0)},0)$ which are cyclically permuted
by the action), as well as the point at infinity.

Thus, each of the cases above has three or four singular points -- all of them orbifold points -- which
give decoupled SCFTs when the curve connecting them goes to infinite area.
In all other cases, the singular points are limited to the origin
and the point at infinity, so there are at most two, again giving decoupled
SCFTs in the infinite area limit.  Assuming that $\Gamma_{U(2)}$ is non-trivial,
the singularity at infinity is an
orbifold, but the singularity at the origin need not be.

In all of these cases, the polynomials
$f$ and $g$  takes a restricted form which must be compatible with the
overall group action. Moreover, we will see that this typically requires a singular
elliptic fibration since $f$ and $g$ must necessarily vanish at the location
of the fixed point.

Let us illustrate this point for cyclic subgroups of $U(2)$. These are
dictated by two relatively prime positive integers $p$ and $q$ with generator
$\omega=\exp(2\pi i/p)$:%
\begin{equation}
\gamma:\left(  s,t\right)  \mapsto(\omega s,\omega^{q}t).
\end{equation}
The minimal resolution of the orbifold singularity is described by a
collection of curves of self-intersection $-n_{1},...,-n_{k}$, where the
sequence also indicates which curves intersect. The values $p$ and $q$ are
dictated by the continued fraction:%
\begin{equation}
\frac{p}{q}=n_{1}-\frac{1}{n_{2}-...\frac{1}{n_{k}}}.
\end{equation}

The specific fractions $p/q$ which can appear in F-theory constructions have
been catalogued in \cite{Heckman:2013pva, Morrison:2016nrt}.
Expanding $f$ and $g$ as polynomials in the variables $s$ and $t$,%
\begin{align}
f  &  =\underset{i,j}{\sum}f_{ij}s^{i}t^{j}\\
g  &  =\underset{i,j}{\sum}g_{ij}s^{i}t^{j},
\end{align}
the group action by $\gamma$ is:%
\begin{align}
f  &  \mapsto\underset{i,j}{\sum}\omega^{i+qj}f_{ij}s^{i}t^{j}=\omega
^{4+4q}\underset{i,j}{\sum}f_{ij}s^{i}t^{j}\\
g  &  \mapsto\underset{i,j}{\sum}\omega^{i+qj}g_{ij}s^{i}t^{j}=\omega
^{6+6q}\underset{i,j}{\sum}g_{ij}s^{i}t^{j},
\end{align}
where in the second equality of each line, we have used the conditions of
lines (\ref{faction}) and (\ref{gaction}). This restricts the available non-zero coefficients:
\begin{align}
f_{ij}  &  \neq0\text{ \ \ only for \ \ }i+qj\equiv4+4q\text{ mod
}p\label{fgen}\\
g_{ij}  &  \neq0\text{ \ \ only for \ \ }i+qj\equiv6+6q\text{ mod }p.
\label{ggen}%
\end{align}
In most cases, this requires both $f$ and $g$ to vanish to some prescribed
order, and we present examples of this type in section \ref{sec:EXAMPLES}. Let us note that
to extract the theory on the tensor branch, we will of course need to perform
further blowups in the base, which will in turn lead to higher order vanishing
for $f$ and $g$. The minimal order of vanishing is generic, but we can also
entertain higher order vanishing for $f$ and $g$. In such cases, we must
perform a resolution of the Calabi--Yau threefold

To illustrate the above, consider the case of an F-theory base given by a
single curve of self-intersection $-3$. In the limit where this curve
collapses to zero size, we have an orbifold singularity $\mathbb{C}%
^{2}/\mathbb{Z}_{3}$, and the polynomials $f$ and $g$ satisfy:%
\begin{align}
f_{ij}  &  \neq0\text{ \ \ only for \ \ }i+j+1\equiv0\text{ mod }3\\
g_{ij}  &  \neq0\text{ \ \ only for \ \ }i+j\equiv0\text{ mod }3,
\end{align}
so to leading order, we have:%
\begin{equation}
f=f_{2,0}s^{2}+f_{1,1}st+f_{0,2}t^{2}+...\text{ \ \ and \ \ }g=g_{0,0}+...
\end{equation}
Following a similar set of steps, we can analyze each case of an orbifold
group action $\Gamma_{U(2)}\subset U(2)$ which appears in the classification
results of \cite{Heckman:2013pva}.

\section{Illustrative Examples \label{sec:EXAMPLES}}

In the previous section we presented a general algorithm for constructing a
large class of 5D\ fixed points. This procedure consists of writing down the
Weierstrass model over a singular base, with the Weierstrass model
coefficients $f$ and $g$ given by suitable $\Gamma_{U(2)}$ equivariant
polynomials. Due to the way we have constructed the model as a canonical
singularity, we are guaranteed to generate at least one 5D\ fixed point of some sort. It
is natural to ask, however, whether we can extract additional details on this
theory, for example, the structure of the 5D\ effective field theory on the
Coulomb branch. Rather than embark on a systematic
classification of all such possibilities, we will mainly focus on some
illustrative examples. Most of the important elements of this analysis can
already be seen for the case of $\Gamma_{U(2)}$ a cyclic group, so we confine
our attention to this case. This already covers all of the non-Higgsable
cluster theories, as well as the \textquotedblleft A-type rigid
theories\textquotedblright\ of \cite{Heckman:2013pva}, namely those without any
complex structure deformations.

\subsection{Non-Higgsable Clusters}

Let us begin by cataloguing the phase structure of the non-Higgsable cluster
theories. Recall that these are given in F-theory by specific collections of
up to three curves, in which the minimal elliptic fibration is always
singular. The collection of curves of self-intersection $-n$
and corresponding 6D gauge algebra are:%
\begin{equation}%
\begin{tabular}
[c]{|l|l|l|l|l|l|l|l|l|l|l|}\hline
Curves & $3$ & $4$ & $5$ & $6$ & $7$ & $8$ & $12$ & $3,2$ & $3,2,2$ &
$2,3,2$\\\hline
$\mathfrak{g}$ & $\mathfrak{su}(3)$ & $\mathfrak{so}(8)$ & $\mathfrak{f}_{4}$
& $\mathfrak{e}_{6}$ & $\mathfrak{e}_{7}$ & $\mathfrak{e}_{7}$ & $e_{8}$ &
$\mathfrak{g}_{2}\times\mathfrak{su}(2)$ & $g_{2}\times\mathfrak{sp}(1)$ &
$\mathfrak{su}(2)\times\mathfrak{so}(7)\times\mathfrak{su}(2)$\\\hline
\end{tabular}
\end{equation}
In the case of the $-7$ curve theory and multiple curve non-Higgsable
clusters, there are also half-hypermultiplet matter fields.

Dimensional reduction on the tensor branch yields a few interesting features.
First of all, for all of the single curve theories, we have just a single simple
gauge group factor, and the number of matter fields is either zero or a single
half hypermultiplet in the fundamental (for the $-7$ curve theory), so we
expect to realize a 5D conformal fixed point on this branch. The resulting
configuration of divisors are, for the simply laced gauge algebras, just a
higher-dimensional analogue of Dynkin diagrams in which the diagram indicates
the intersection of Hirzebruch surfaces. See Appendix A for details.

\begin{figure}
\begin{center}
\includegraphics[scale=1.2]{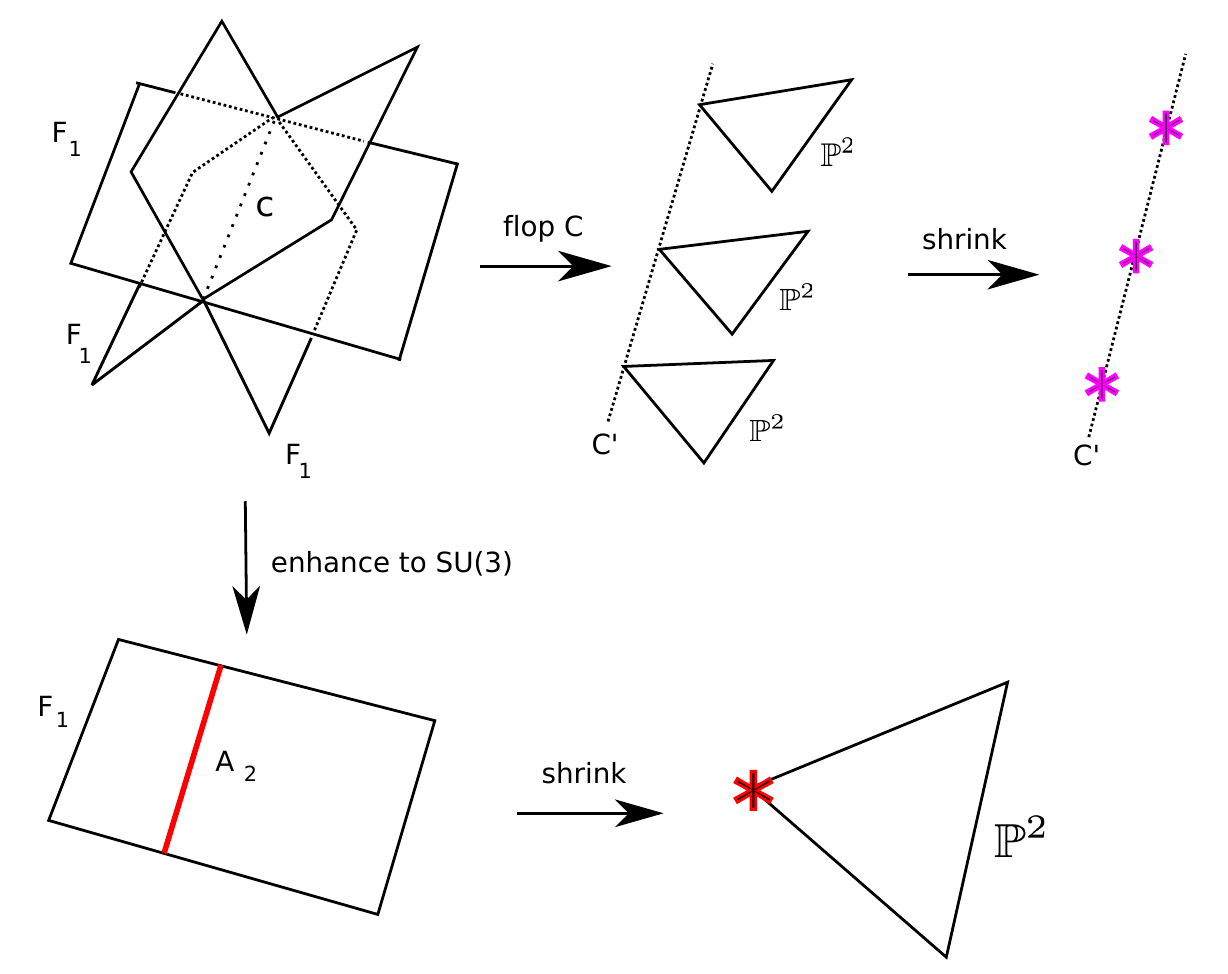}
\end{center}
\caption{Geometry of the $-3$ theory. \textsc{upper left}: Reduction of the tensor branch over $S^1$; \textsc{upper center}: flop phase transition; \textsc{upper right}: reduction of the 6D SCFT over $S^1$; \textsc{lower left}: gauge symmetry enhanced to $SU(3)$; \textsc{lower right}: strong coupling limit of $SU(3)$ theory.  In the 5D limit, $C'$ and $\mathbb{P}^2$ decompactify.}\label{fig:minus3}
\end{figure}

Let us discuss the physics of this reduction in more detail for one example, the case of the $-3$ curve. The resulting geometry is depicted in Figure \ref{fig:minus3}. By reducing on the circle the tensor branch of this theory, we obtain a collection of $\mathbb{F}_1$ Hirzebruch surfaces which intersect giving rise to a Kodaira type $IV$ fiber. In Figure \ref{fig:minus3} we have indicated the curve which we can flop by $C$. It is a rational curve with an $\mathcal{O}(-1)\oplus \mathcal{O}(-1)$ normal bundle. Flopping it we obtain a curve $C'$ with three $\mathbb{P}^2$ surfaces intersecting it at a point. Shrinking these surfaces down to zero size we obtain three 5D SCFTs corresponding to $\mathbb{C}^3/\mathbb{Z}_3$ orbifold points. The remaining curve has the same area as the nearby elliptic curves,
so in
the limit $R_{S^1}\to 0$, the curve $C'$ grows to infinite size and the three $\mathbb{C}^3/\mathbb{Z}_3$ theories decouple.

In this case, the $S^1$ reduction of the 6D tensor branch also flows to a fixed point, corresponding to the pure $SU(3)$ gauge group without matter (the $U(1)$ vector multiplet corresponding to the dimensional reduction of the 6D tensor
multiplet decouples).
This is  illustrated in the lower portion of Figure \ref{fig:minus3}.  One first shrinks two of the $\mathbb{F}_1$ surfaces to the common
curve of intersection, where they form a curve of $A_2$ singularities.
To take that gauge theory to strong coupling, we shrink the area of
the curve of singularities, leaving a single $\mathbb{P}^2$ containing
a single conformal point (the strongly coupled $SU(3)$ theory).

This example is interesting because it illustrates how, even in a  simple situation, non-trivial 5D fixed points can occur in different chambers of the extended K\"ahler cone. The fact that we obtain a 5D SCFT from the phase corresponding to the $S^1$ reduction of the tensor branch has to be regarded as a coincidence, though. The actual reduction of the 6D SCFT on $S^1$ is given by the three $\mathbb{C}^3 / \mathbb{Z}_3$ theories.

\begin{figure}
\begin{center}
\includegraphics[scale=1.1]{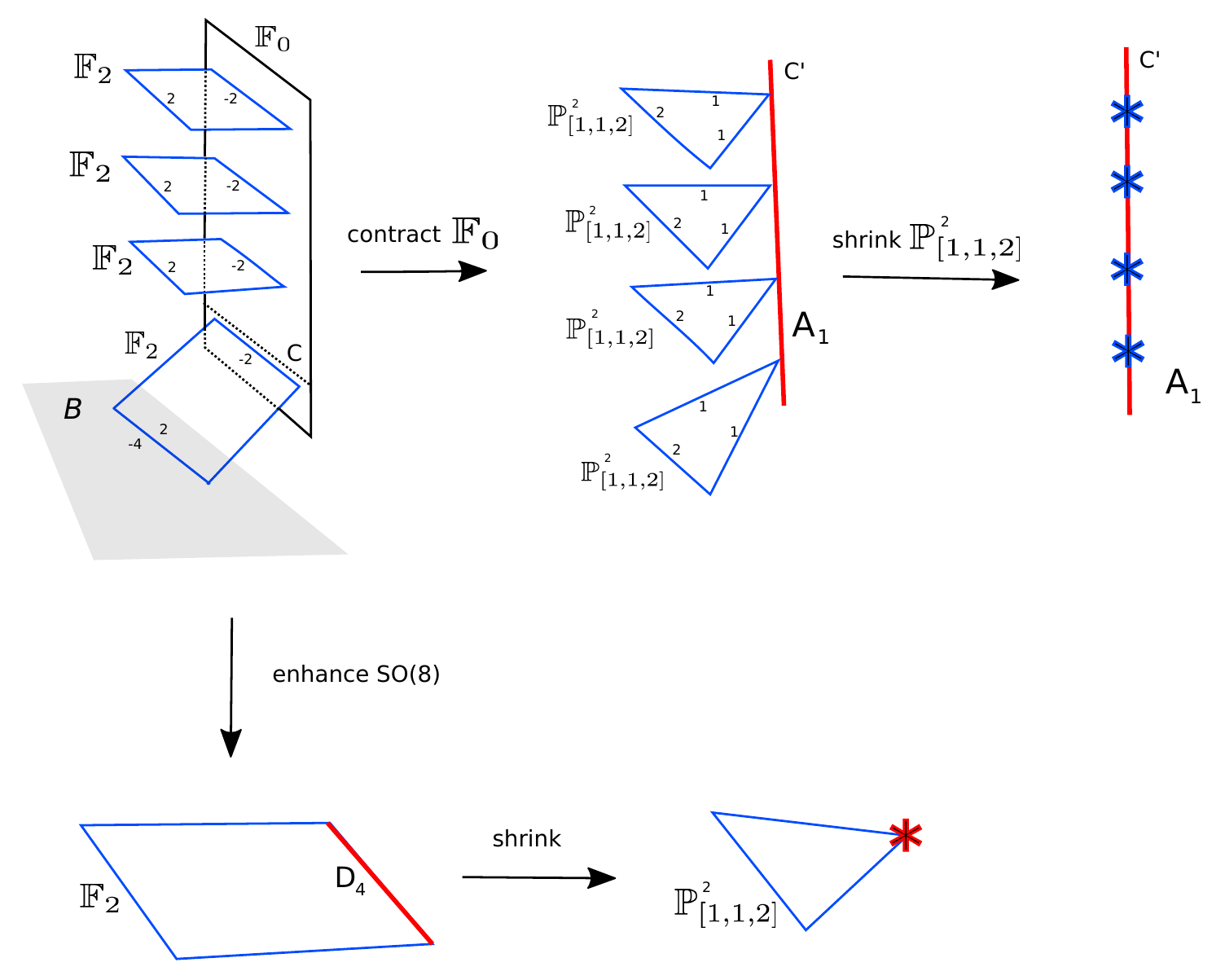}
\end{center}
\caption{Geometry of the $-4$ theory. \textsc{upper left}: Reduction of the tensor branch over $S^1$; \textsc{upper center}: flop phase transition; \textsc{upper right}: reduction of the 6D SCFT over $S^1$\textsc{lower left}: gauge symmetry enhanced to $SO(8)$; \textsc{lower right}: strong coupling limit of $SO(8)$ theory.
In the 5D limit, the $A_1$ locus and $\mathbb{P}^2_{[1,1,2]}$ decompactify.}\label{fig:D4_nhc}
\end{figure}

As a second example we consider
the case of the $-4$ curve. The resulting geometry is depicted in Figure \ref{fig:D4_nhc}. By reducing on the circle the tensor branch of this theory, we obtain an $\mathbb{F}_0$ Hirzebruch surface meeting four $\mathbb{F}_2$ Hirzebruch surfaces along fibers of one of the rulings of $\mathbb{F}_0$.  The intersection pattern gives
rise to a Kodaira type $I_0^*$ fiber. This time, instead of flopping a curve we contract a divisor to a curve, in one of two different ways.  If we contract the $\mathbb{F}_0$ along the ruling which includes the intersection curves with the $\mathbb{F}_2$ surfaces, we obtain
a curve of $SU(2)$ singularities with four $\mathbb{P}^2_{[1,1,2]}$ surfaces intersecting it at a point. Shrinking these surfaces down to zero size we obtain four 5D SCFTs corresponding to $\mathbb{C}^3/\mathbb{Z}_4$ orbifold points with group action specified by $(\frac14,\frac14,\frac12)$. The corresponding curve of $A_1$ singularities gives an $SU(2)$ gauge group with gauge coupling $g^2_{SU(2)}\sim 1/\text{vol}(C)$ which is also proportional to $R_{S^1}$. In the limit $R_{S^1}\to 0$, the curve $C$ grows to infinite size and the four $\mathbb{C}^3/\mathbb{Z}_4$ theories decouple. These models have an $SU(2)$ flavor symmetry.

In this case, the $S^1$ reduction of the 6D tensor branch also flows to a fixed point, corresponding to the pure $SO(8)$ gauge group without matter.
That is illustrated in the lower portion of Figure \ref{fig:D4_nhc}.
One first shrinks the $\mathbb{F}_0$ along its other ruling together with three of the  $\mathbb{F}_2$ surfaces to a
curve of $D_4$ singularities.
To take that gauge theory to strong coupling, we shrink the area of
the curve of singularities, leaving a single $\mathbb{P}^2_{[1,1,2]}$ containing
a single conformal point (the strongly coupled $SO(8)$ theory).

For the multiple curve theories, however, we do not expect to realize a
conformal fixed point in the chamber corresponding to the $S^1$ reduction of the  moduli space. This again follows from
the criterion put forward in \cite{Intriligator:1997pq}, because we always have a product gauge
group with bifundamental matter. To reach a conformal fixed point for these
geometries, we must perform a flop transition to another chamber of moduli
space, namely that described by the orbifold procedure outlined above.

We can carry out the analysis of section \ref{sec:REDUCE} for each of
these examples quite explicitly.  In Table \ref{tab:fg}, for each $p/q$
corresponding to a non-Higgsable cluster, we describe the finite group
action on the variables $s$, $t$, $x$, $y$ and functions $f$, $g$ which appear in
the corresponding Weierstrass equation, and we also give the lowest
order terms in $f$ and $g$.  This data then determines the
5D fixed points after $S^1$ reduction.

\begin{table}[h]
\begin{center}
{\renewcommand{\arraystretch}{1.5}
\begin{tabular}{cccc}
$p/q$ &  $(s,t,x,y;f,g)$ & $f$ & $g$ \\ \hline
$2$ & $(\frac12,\frac12,0,0;0,0)$ & $f_0$ & $g_0$ \\
$3$ & $(\frac13,\frac13,\frac13,0;\frac23,0)$ & $f_0s^2+f_1st+f_2t^2$ &$g_0$\\
$4$ & $(\frac14,\frac14,0,\frac12;0,0)$ & $f_0$ & $g_0$\\
$5$ & $(\frac15,\frac15,\frac45,\frac15;\frac35,\frac25)$
  & $f_0s^3+f_1s^2t+f_2st^2+f_3t^3$ & $g_0s^2+g_1st+g_2t^2$ \\
$6$ & $(\frac16,\frac16,\frac23,0;\frac13,0)$
  & $f_0s^2+f_1st+f_2s^2$ & $g_0$ \\
$7$ & $(\frac17,\frac17,\frac47,\frac67;\frac17,\frac57)$
  & $f_0s+f_1t$ & $\sum_{j=0}^5g_js^{5-j}t^j$ \\
$8$ & $(\frac18,\frac18,\frac12,\frac34;0,\frac12)$
  & $f_0$ &
  $\sum_{j=0}^4g_js^{4-j}t^j$\\
$12$ & $(\frac1{12},\frac1{12},\frac13,\frac12;\frac23,0)$
  & $f_0=\sum_{j=0}^8f_js^{8-j}t^j$ & $g_0$ \\
$5/2$ & $(\frac15,\frac25,\frac15,\frac45;\frac25,\frac35)$
  & $f_0s^2+f_1t$ & $g_0s^3+g_1st+g_2t^4$\\
$7/3$ & $(\frac17,\frac37,\frac17,\frac57;\frac27,\frac37)$
  & $f_0s^2+
f_2t^3 $ & $g_0s^3+g_1t$ \\
$8/5$ & $(\frac18,\frac58,\frac12,\frac14;0,\frac12)$
  & $f_0$ & $g_0s^4+g_1s^2t^2+g_2t^4$ \\
\end{tabular}
}
\caption{Weierstrass coefficients.  All $f_j$ and $g_j$ are $\Gamma_{U(2)}$-invariant
functions.}\label{tab:fg}
\end{center}
\end{table}

In order to see the geometry of the fixed points, we need
 to determine the fixed point set of the group action,
and what subgroup stabilizes each fixed point.  This information is
tabulated in Table~\ref{tab:loci}.  The origin is always
fixed by the entire group, but if $g_0$ is constant, the hypersurface
does not pass through the origin; in that case, we have written ``no''
in the non-orbifold column.  The orbifold points are specified by
their group actions.

\begin{table}[h]
\begin{center}
{\renewcommand{\arraystretch}{1.5}
\begin{tabular}{ccccccc}
$p/q$ &  $(stxy;fg)$ & codim $2$ & orbifold points & non-orbifold point\\ \hline
$2$ & $(\frac12,\frac12,0,0;0,0)$  & $A_1$ & none & no\\
$3$ & $(\frac13,\frac13,\frac13,0;\frac23,0)$
  & none & $3\times (\frac13,\frac13,\frac13)$ & no\\
$4$ & $(\frac14,\frac14,0,\frac12;0,0)$
  & $A_1$ & $4\times (\frac14,\frac14,\frac12)$ &no\\
$5$ & $(\frac15,\frac15,\frac45,\frac15;\frac35,\frac25)$
  & none & $(\frac15,\frac15,\frac35)$ & yes \\
$6$ & $(\frac16,\frac16,\frac23,0;\frac13,0)$
  & $A_1$ & $3\times (\frac16,\frac16,\frac23)$ & no\\
$7$ & $(\frac17,\frac17,\frac47,\frac67;\frac17,\frac57)$
  & none & $(\frac17,\frac17,\frac57)$ & yes\\
$8$ & $(\frac18,\frac18,\frac12,\frac34;0,\frac12)$
  & $A_1$ & $(\frac14,\frac14,\frac12)$; $2\times (\frac18,\frac18,\frac34)$ & no\\
$12$ & $(\frac1{12},\frac1{12},\frac13,\frac12;\frac23,0)$
  & $A_1$ &
$(\frac14,\frac14,\frac12)$;
$(\frac16,\frac16,\frac23)$;
$(\frac1{12},\frac1{12},\frac56)$& no\\
$5/2$ & $(\frac15,\frac25,\frac15,\frac45;\frac25,\frac35)$
  &none&$(\frac15,\frac25,\frac25)$&yes\\
$7/3$ & $(\frac17,\frac37,\frac17,\frac57;\frac27,\frac37)$
  &none&$(\frac17,\frac37,\frac37)$ & yes\\
$8/5$ & $(\frac18,\frac58,\frac12,\frac14;0,\frac12)$
  &$A_1$&$2\times (\frac14,\frac14,\frac12)$; $(\frac18,\frac58,\frac14)$&no\\
\end{tabular}
}
\end{center}
\caption{Singularity loci.} \label{tab:loci}
\end{table}

\subsection{Rigid A-type Theories}

Consider next the Rigid A-type theories of reference \cite{Heckman:2013pva}.
These are defined by considering a base $B$ with collapsing curves intersecting as:%
\begin{equation}
n_{1},...,n_{k}.
\end{equation}
We then perform the minimal resolutions necessary to place all elliptic fibers
in Kodaira-Tate form. We denote the Hirzebruch-Jung continued fraction by
$p/q$. These theories have no continuous flavor symmetries in six dimensions.
Consequently, any flavor symmetries obtained upon reduction to five dimensions
should be viewed as emergent in the infrared.

There are at least two disconnected components to the 5D\ SCFT, and
there may be three or four.  To determine which case occurs, we
follow the analysis in section \ref{sec:REDUCE} and see that it is
determined by the knowledge of which power of the determinant vanishes.

In Appendix A of \cite{Morrison:2016nrt}, the rigid theories are listed
and their determinants are computed.  The cases of interest here appear
in block diagonals of the tables in that paper, and in particular, the
analysis there shows that there are infinite families of examples
for each of the cases analyzed in section \ref{sec:REDUCE}.  That is,
there are infinite families of examples with four orbifold points,
or with three orbifold points of the same type, and so on.  What
changes is the codimension 2 singular locus, which can give a (flavor)
symmetry of arbitrarily large rank.

For example, $p/q = 4N/(2N-1)$ corresponds to the
data
\begin{equation}
(s,t,x,y;f,g) = \left(\frac1{4N},\frac{2N-1}{4N},0,\frac12;0,0\right)
\end{equation}
 and there are four orbifold points
of type $(\frac1{4N},\frac{2N-1}{4N},\frac12)$ with a codimension two locus supporting
an $A_{2N-1}$ singularity.
When the base is fully resolved, it corresponds to
$4141\cdots14$.

\subsection{M5-Brane Probe Theories}

It is also of interest to consider 6D\ SCFTs with a non-trivial Higgs branch.
A canonical class of examples are provided by M5-branes probing an
ADE\ singularity, and M5-branes probing a Ho\v{r}ava-Witten $E_{8}$ wall, or
combinations thereof.

\subsubsection{Probes of an ADE\ Singularity} \label{sec531}

Consider first the case of M5-branes probing an ADE\ singularity. The F-theory
realization of these 6D\ SCFTs is straightforward to realize in terms of a
pair of colliding singularities, each associated with an algebra of type
$\mathfrak{g}_{ADE}$ which intersect at the singular point of the geometry
$\mathbb{C}^{2}/\mathbb{Z}_{k}$. Minimal resolution of the orbifold in the
base yields a chain of $-2$ curves, and the presence of the colliding
singularities gives an additional enhancement in the singularity type over
each $-2$ curve. The partial tensor branch is then given by:%
\begin{equation}
\lbrack\mathfrak{g}]\overset{\mathfrak{g}}{2},...,\overset{\mathfrak{g}%
}{2}[\mathfrak{g}]. \label{partialnow}%
\end{equation}
In the M5-brane picture, this corresponds to seperating the branes along the
$\mathbb{R}_{\bot}$ factor of $\mathbb{R}_{\bot}\times\mathbb{C}%
^{2}/ \Gamma_{ADE}$. Further blowups between each such collision are
required to place all elliptic fibers in Kodaira-Tate form. Returning to the
partial tensor branch of line (\ref{partialnow}), we can read off the
reduction to five dimensions. It is given by a generalized 5D quiver, with
gauge algebras $\mathfrak{g}_{ADE}$, and 5D conformal matter. This 5D conformal
matter is the CFT associated with compactification of 6D conformal
matter and as such, the analysis of section \ref{sec:REDUCE} guarantees that we
will indeed reach a fixed point. On the Coulomb branch, this system is, after taking
an appropriate flop transition described by the affine quiver gauge theory obtained from
D4-branes probing an  ADE singularity. Indeed, we note that when we have more than one gauge group factor, the
argument of \cite{Intriligator:1997pq} applies, and we do not expect a 5D fixed point in the
chamber of moduli space where the quiver gauge theory description is valid.
If we go to the full 6D tensor branch and then reduce, we encounter a similar issue.

To reach a 5D\ fixed point, we would need to perform a sequence of flop
transitions, and one region of moduli space where we are guaranteed to find
such a fixed point is in circle reduction of the 6D\ fixed point. Indeed, the
F-theory model for this case is also straightforward to engineer. To see why,
consider first the model for a single component of the discriminant locus of
type $\mathfrak{g}_{ADE}$. We can parameterize this in terms of the local
equation:%
\begin{equation}
y^{2}=x^{3}+f(s)x+g(s),
\end{equation}
for a single holomorphic coordinate $s$ of $\mathbb{C}$. In all but the
$I_{n}$ fiber case, the leading order behavior of this singularity takes the
form:%
\begin{equation}
y^{2}=x^{3}+s^{a}x+s^{b},
\end{equation}
for some suitable choice of $a$ and $b$. To realize a collision in
$\mathbb{C}^{2}$, we then have (see e.g. \cite{DelZotto:2014hpa, Heckman:2014qba}):
\begin{equation}
y^{2}=x^{3}+(st)^{a}x+(st)^{b}. \label{collider}%
\end{equation}

Importantly, we note that the further quotient by $(s,t)\rightarrow(\omega
s,\omega^{-1}t)$ imposes no additional restrictions on the form of line
(\ref{collider}), so we conclude that $a$ and $b$ (as dictated by the choice
of gauge algebra) remain the same for this model.

Note also that in this case,
the \textquotedblleft patch at infinity\textquotedblright\ with $y=1$ does not
actually contribute a 5D\ SCFT. The reason is that the orbifold locus is
locally given by $\mathbb{C}\times\mathbb{C}^{2}/\Gamma_{ADE}$, and so there
are no collapsing divisors in this region of the geometry. Instead, all of the
collapsing divisors are concentrated in the patch described by line
(\ref{collider}).

As a concrete example, we see that the form of colliding $E_{8}$
singularities, namely a collision of two type $II^{\ast}$ fibers, is:%
\begin{equation}
y^{2}=x^{3}+(st)^{4}x+(st)^{5}.
\end{equation}
We produce a 5D generalized quiver with $E_8$ gauge group factors and
$(E_8,E_8)$ conformal matter by performing a $\mathbb{Z}_k$ quotient
on the base. Though it would be interesting to perform a similar
analysis of the fully resolved geometry (akin to what
we did for the non-Higgsable cluster theories) and to then collapse divisors to reach a canonical
singularity, this will of course be much more involved due
to the large number of additional compact cycles in this case. We leave this interesting issue for future work.

\subsubsection{Probes of an $E_{8}$ Wall}

Consider next the case of M5-branes next to an $E_{8}$ nine-brane. The F-theory model
has a base:
\begin{equation}
\lbrack E_{8}]\underset{k}{\underbrace{1,2,...,2}}, \label{instantons}%
\end{equation}
where the $E_{8}$ flavor symmetry is only manifest in the limit where all
curves collapse to zero size. The associated Weierstrass model is:
\begin{equation}
y^{2}=x^{3}+g_{k}(s)t^{5},
\end{equation}
where $g_{k}(s)$ is a degree $k$ polynomial in $s$.

The dimensional reduction of this model to five dimensions has already been
determined in the literature. It is given by an $Sp(k)$ gauge theory with
$N=7$ hypermultiplets in the fundamental representation. In the limit where
the gauge theory passes to strong coupling, the flavor symmetry enhances from
$SO(14)$ to $E_{8}$.

The geometry of the $k=1$ case is already quite interesting. The local
geometry for this case is a del Pezzo nine surface. Flopping the zero section,
we reach the standard description in terms of a local $dP_{8}$ which can
contract to zero size. In the case of $k>1$, this flop also converts the local
surface associated with the $-2$ curve to another $dP_{9}$. One can see this
since the blowdown of the $-1$ curve converts the leftmost $-2$ curve to a
$-1$ curve. This in turn means we get another local $dP_{9}$ geometry.
Continuing in this fashion, we obtain a chain of intersecting $dP_{8}$
surfaces, all of which are collapsing to zero size.

We can also consider a non-trivial fiber enhancement over the curves of line
(\ref{instantons}). This is interpreted as small instantons probing an
ADE\ singularity \cite{Aspinwall:1997ye, DelZotto:2014hpa , Heckman:2015bfa}. In this case,
the partial tensor branch is not expected to realize a 5D\ SCFT upon circle
reduction. We can, however, again take a flopped phase of the geometry, i.e.,
keep all curves of the base at small size when we pass to five dimensions. In
this case, we again expect to realize a 5D\ SCFT.

\section{Conclusions \label{sec:CONC}}

The classification of 6D\ SCFTs via F-theory provides a starting point for the
construction and study of lower-dimensional SCFTs. In this paper we have
applied these general considerations in the study of 5D\ SCFTs. Starting from
6D\ SCFTs realized via F-theory on an elliptically fibered Calabi--Yau
threefold, we have shown how further reduction on a circle leads to a rich
phase structure for 5D theories, as realized by M-theory compactified on the
same Calabi--Yau. In particular, we have seen that the reduction of
a 6D $\mathcal{N} = (1,0)$ SCFT to five dimensions yields a 5D\ SCFT, and moreover, the reduction
of the tensor branch deformation of a 6D\ SCFT typically does not yield a
5D\ SCFT. In the Calabi--Yau geometry, the two phases are connected by a
sequence of flop transitions, namely a trajectory in the extended K\"ahler cone.
The existence of these two phases provides a concrete way to pass from one
phase to the other, namely, by a flow through moduli space. By elucidating the
structure of the 5D conformal fixed points, we have shown in particular how 5D
quiver gauge theories can be connected to a class of geometrically realized
fixed points. In the remainder of this section we discuss some avenues of
future investigation.

One of the important uses of a 5D gauge theory analysis is the potential to
explicitly compute the structure of an associated supersymmetric index. Now,
even though we have argued that one must flop to another chamber of moduli
space to actually realize the fixed point, the sense in which this object
transforms under flops should be well controlled. In this sense, gauge theory
methods for calculating such quantities should have an interpretation in terms
of a superconformal index. This is indeed the philosophy adopted in much of
the literature on 5D SCFTs (see e.g. \cite{Bergman:2013koa, Bergman:2013ala, Papageorgakis:2016cej}),
though with the explicit geometry now in hand, one can in principle check these claims by direct
calculation of topological string amplitudes on the Calabi--Yau in the conformal chamber, perhaps
along the lines of \cite{Lockhart:2012vp}.

Now that we have constructed a broad class of new 5D SCFTs, it is natural
to ask whether some of these also yield holographic duals, perhaps along the lines of
\cite{Apruzzi:2014qva, Kim:2015hya, DHoker:2016ujz, DHoker:2016ysh}. Circle reduction of
$AdS_7$ vacua does not yield $AdS_6$ vacua, which is in accord with the phase structure observed in this work.
We have also seen, however, that flop transitions often yield a 5D fixed point. It would be interesting to understand
this holographically.

Perhaps more ambitiously, one might hope to also classify all interacting
5D\ SCFTs. From a geometric standpoint, this would require understanding all
local Calabi--Yau models with divisors which can simultaneously contract to a
point. In particular, it would be interesting to determine whether some
generalization of the numerical invariants used in the classification of
6D\ SCFTs can be obtained for this class of geometries as well. Let us note
that from a physical perspective, one might be tempted to conjecture that all
5D\ SCFTs are obtained from some deformation of a 6D\ SCFT on a circle. This
looks difficult to arrange in all cases, since, for example, supersymmetric
orbifolds of the form $\mathbb{C}^{3}/\Gamma_{SU(3)}$ for $\Gamma_{SU(3)}$ a
finite subgroup of $SU(3)$ do not have a clear embedding in an elliptically
fibered Calabi--Yau threefold of the sort used to engineer 6D\ SCFTs via
F-theory. Either establishing a firm counterexample, or developing a clear
method of embedding 5D\ SCFTs in 6D theories would be most instructive.

\section*{Acknowledgements}

MDZ and JJH thank C. Vafa for helpful discussions and collaboration at an early
stage of this work. We thank F. Apruzzi, T. Dumitrescu, M. Esole, N. Seiberg and W. Taylor
for helpful discussions.
We thank the 2016 Summer Workshop at the Simons Center for Geometry and Physics
for hospitality during this work, as well as the Aspen Center for Physics
Winter Conference on Superconformal Field Theories in $d \geq 4$, NSF grant PHY-1066293,
during the final stages of this work.
The work of MDZ is supported by NSF grant PHY-1067976 and by SCGP. The work of JJH
is supported by NSF CAREER grant PHY-1452037. JJH also acknowledges support
from the Bahnson Fund at UNC Chapel Hill. The work of DRM is
supported in part by NSF grants PHY-1307513 and PHY-1620842 and by the
Centre National de la Recherche Scientifique (France).


\newpage

\appendix

\section{Rank One NHCs on a Circle}

In this Appendix we provide additional details on the resolution of the rank
one non-Higgsable cluster theories.\ Recall that for these theories, both the
6D tensor branch and conformal fixed point yield 5D\ SCFTs, which are, as
usual, connected by a flop transition. For the other non-Higgsable cluster
theories, we have at least two gauge group factors, so the argument of
\cite{Intriligator:1997pq} already tells us that we will not be able to reach
a 5D\ SCFT by reducing the tensor branch. Rather, we must perform a flop
transition to reach a 5D\ SCFT. We proceed by analyzing the single $-n$ curve
theories, splitting up our analysis into the cases of a simply laced Lie
algebra with no matter, and then all other cases.

\subsection{$n=3,4,6,8,12$ Theories}

Consider, then, a single $-n$ curve theory, and assume that the minimal fiber
type leads to a simply laced Lie algebra with no enhancements over the base
curve. The local Calabi--Yau geometry is described by a curve of ADE-type
singularities, and the resolution of these singularities is well-known:
Including the elliptic fiber class, we get a collection of $-2$ curves which
intersect according to the affine extension of the Dynkin diagram. Roughly
speaking, we need to understand how these $-2$ curves fiber over the base $-n$
curve to produce a collection of compact divisors in our non-compact
Calabi--Yau threefold.

Our main claim is that the collection of compact divisors are Hirzebruch
surfaces which intersect according to the affine Dynkin diagram. Recall that
for a Hirzebruch surface of degree $k$, we have a $\mathbb{P}^{1}$ fibered over a base
$\mathbb{P}^{1}$, and the degree of this fibration is $k$. Introducing a base class $b$
and fiber class $f$, we have the intersection numbers:%
\begin{equation}
b\cdot b=-k\text{, \ \ }b\cdot f=1\text{, \ \ }f\cdot f=0\text{.}%
\end{equation}
There are actually two zero sections. One is given by $b$, and the other is
given by $b+kf$. Note that this class has self-intersection:%
\begin{equation}
(b+kf)\cdot(b+kf)=-k+2k=k\text{.}%
\end{equation}

Let us now establish that we indeed have a configuration of intersecting
Hirzebruch surfaces. To understand this, consider the $-n$ curve of the base.
Since we can fully resolve the singular fiber, the local geometry for this
curve is given by the total space $\mathcal{O}(-n)+\mathcal{O}(n-2)\rightarrow
\mathbb{P}^{1}$. What this means is that the affine node of the Dynkin diagram
fibers over this curve as a bundle of degree $n-2$. This is simply the
geometry of a Hirzebruch surface of degree $n-2$, which we denote by
$\mathbb{F}_{n-2}$. Going to the other zero section of this divisor, the local
geometry is now given by $\mathcal{O}(-(n-2))+\mathcal{O}((n-2)-2)\rightarrow
\mathbb{P}^{1}$. Indeed, the neighboring node of the Dynkin diagram also
defines a $\mathbb{P}^{1}$, and it fibers over a $\mathbb{P}^{1}$ as well.
Said differently, we see that the neighboring node defines a degree $n - 4$
Hirzebruch surface. The surfaces intersect along a $\mathbb{P}^{1}$ which we
denote by $C_{n-2,n-4}$:
\begin{equation}
\mathbb{F}_{n-2} \cdot_{CY} \mathbb{F}_{n - 4} = C_{n-2,n-4}.
\end{equation}
The self-intersection of this curve in each of the Hirzebruch surfaces is:
\begin{equation}
C_{n-2,n-4} \cdot_{\mathbb{F}_{n-2}} C_{n-2,n-4} = n - 2 \,\,\,
\text{and}\,\,\, C_{n-2,n-4} \cdot_{\mathbb{F}_{n-4}} C_{n-2,n-4}=
-(n-4).
\end{equation}

\begin{figure}
\begin{center}
\includegraphics[scale=1.1]{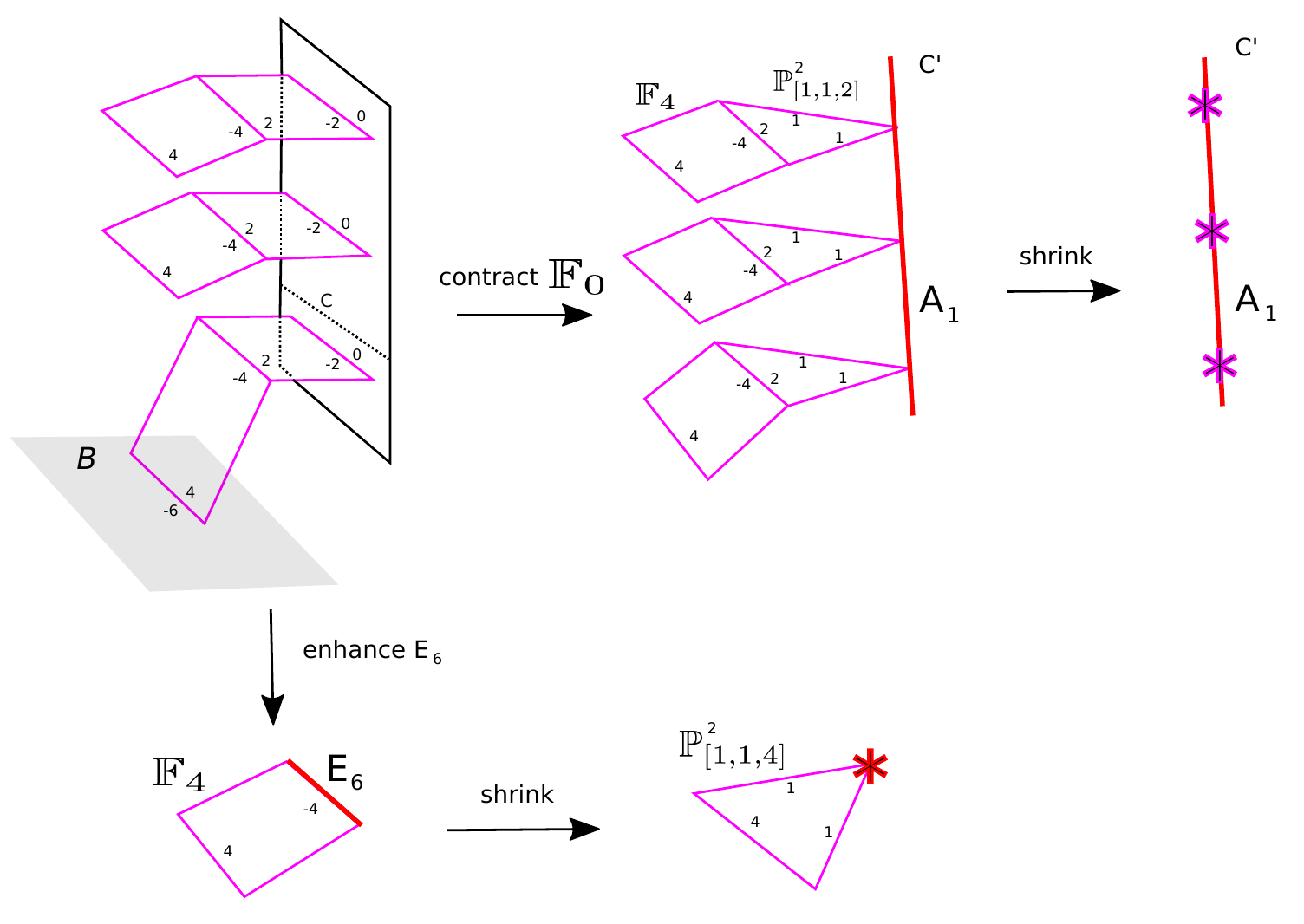}
\end{center}
\caption{Geometry of the $-6$ theory. The base of the elliptic fibration is the noncompact surface $B$. For each $\mathbb{P}^1$ with a non-trivial self-intersection number inside a given surface, the latter is indicated within the corresponding surface.  In the 5D limit, the $A_1$ locus and $\mathbb{P}^2_{[1,1,4]}$ decompactify.} \label{minus6}
\end{figure}

\begin{figure}
\begin{center}
\begin{align}
n  &  =3:\qquad\qquad\begin{gathered}
\xymatrix{&*++[o][F-]{\mathbb{F}_{1}}\ar@{-}[dr]\ar@{-}[dl]&\\
*++[o][F-]{\mathbb{F}_{1}}\ar@{-}[rr]&&*++[o][F-]{\mathbb{F}_{1}}\\
}
\end{gathered}
\nonumber\\
\nonumber\\
n  &  =4:\qquad\qquad\begin{gathered}
\xymatrix{
& *++[o][F-]{\mathbb{F}_{2}} \ar@{-}[d]& \\
*++[o][F-]{\mathbb{F}_{2}} \ar@{-}[r]& *++[o][F-]{\mathbb{F}_{0}} \ar@{-}[r] \ar@{-}[d]& *++[o][F-]{\mathbb{F}_{2}}\\
& *++[o][F-]{\mathbb{F}_{2}} &\\
}
\end{gathered}
\nonumber\\
\nonumber\\
n  &  =6:\qquad\qquad\begin{gathered}
\xymatrix{
&  & *++[o][F-]{\mathbb{F}_{4}} \ar@{-}[d]&  & \\
&  & *++[o][F-]{\mathbb{F}_{2}} \ar@{-}[d]&  & \\
*++[o][F-]{\mathbb{F}_{4}} \ar@{-}[r]& *++[o][F-]{\mathbb{F}_{2}} \ar@{-}[r]& *++[o][F-]{\mathbb{F}_{0}} \ar@{-}[r]& *++[o][F-]{\mathbb{F}_{2}}\ar@{-}[r] &
*++[o][F-]{\mathbb{F}_{4}}
}
\end{gathered}
\nonumber\\
\nonumber\\
n  &  =8:\qquad\qquad\begin{gathered}
\xymatrix{
&  &  & *++[o][F-]{\mathbb{F}_{2}} \ar@{-}[d]&  &  & \\
*++[o][F-]{\mathbb{F}_{6}} \ar@{-}[r]& *++[o][F-]{\mathbb{F}_{4}} \ar@{-}[r]& *++[o][F-]{\mathbb{F}_{2}} \ar@{-}[r]& *++[o][F-]{\mathbb{F}_{0}}\ar@{-}[r] &
*++[o][F-]{\mathbb{F}_{2}} \ar@{-}[r]& *++[o][F-]{\mathbb{F}_{4}} \ar@{-}[r]& *++[o][F-]{\mathbb{F}_{6}}%
}
\end{gathered}
\nonumber\\
\nonumber\\
n  &  =12:\qquad\qquad\begin{gathered}
\xymatrix{
&  &  &  &  & *++[o][F-]{\mathbb{F}_{2}} \ar@{-}[d]&  & \\
*++[o][F-]{\mathbb{F}_{10}} \ar@{-}[r]& *++[o][F-]{\mathbb{F}_{8}} \ar@{-}[r]& *++[o][F-]{\mathbb{F}_{6}} \ar@{-}[r]& *++[o][F-]{\mathbb{F}_{4}} \ar@{-}[r]&
*++[o][F-]{\mathbb{F}_{2}} \ar@{-}[r]& *++[o][F-]{\mathbb{F}_{0}} \ar@{-}[r]& *++[o][F-]{\mathbb{F}_{2}} \ar@{-}[r]& *++[o][F-]{\mathbb{F}_{4}}%
}
\end{gathered}\nonumber
\end{align}
\end{center}
\caption{Schematic structure of the geometries of certain NHCs as Dynkin graphs: the nodes correspond to surfaces while the links correspond to intersections.}\label{DNKN}
\end{figure}
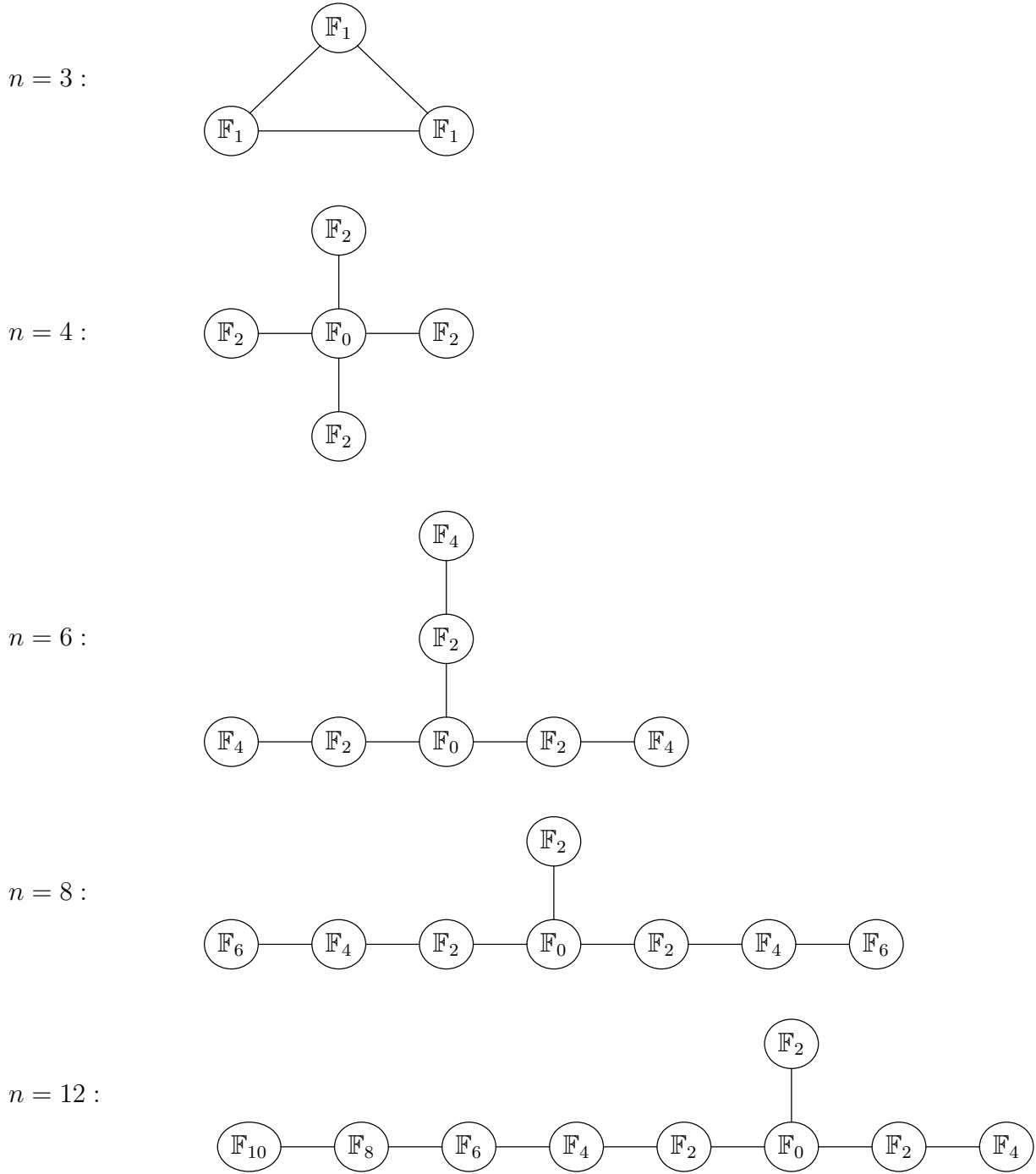

Continuing in this fashion, we see that we build up a collection of Hirzebruch
surfaces, all intersecting according to the affine Dynkin diagram. In the upper left corner of Figure \ref{minus6} we depict the $n=6$
example.\footnote{The remainder of the Figure illustrates how to obtain two different 5D SCFTs from this starting point.}
We list
all these configurations explicitly in Figure \ref{DNKN}%

We remark that in all but the $n=3$ case, there is a \textquotedblleft
middle\textquotedblright\ $\mathbb{F}_{0}$ surface which intersects three or
more additional surfaces. Each of these intersections defines a curve in the
$\mathbb{F}_{0}$ which are homologous, and do not intersect.

Reaching a conformal fixed point now proceeds by first decompactifying the
elliptic curve class, i.e., by decompactifying the Hirzebruch surface
associated with the affine node of the Dynkin diagram. The transition to the
conformal fixed point now proceeds in stages: Collapsing the $\mathbb{F}_{0}$
or $\mathbb{F}_{1}$ causes the neighboring surfaces to become weighted
projective spaces, which can then collapse to zero size. The collapse of these
surfaces causes their neighbors to contract to weighted projective spaces as
well. This process continues until all surfaces have collapsed to zero size.

\subsection{$n=5$ Theory}

Let us now turn to the $-5$ curve theory. Here, the Weierstrass model is (see e.g. \cite{Bershadsky:1996nh}):
\begin{equation}
y^{2}=x^{3}+f_{3}(s)t^{3}x+g_{2}(s)t^{4},
\end{equation}
where $s$ is a local coordinate on the base $\mathbb{P}^{1}$ and $t$ is a
coordinate in the normal directions. The polynomials $f_{3}(s)$ and $g_{2}(s)$
have respective degrees three and two in the variable $s$. The operating
assumption is that $g_{2}(s)$ is generic in the sense that its two roots are
at distinct points. This model realizes a non-split $IV^{\ast}$ fiber, namely
one in which some of the two-cycles of the fiber are identified as we undergo
monodromy in the $s$-plane. Indeed, this monodromy leads to an outer
automorphism of the $\mathfrak{e}_{6}$ algebra to an $\mathfrak{f}_{4}$
algebra in the 6D theory.

Now, from the analysis of reference \cite{Morrison:2012np}, we know that this model
has no localized matter. This in turn means that each
$\mathbb{P}^{1}$ of the degenerate elliptic fiber will fiber over the base,
producing a collection of Hirzebruch surfaces.\footnote{Owing to monodromy in the elliptic fiber,
some of the surfaces are actually a double cover of a $\mathbb{P}^1$ bundle over a $\mathbb{P}^1$. Note, however,
that this double cover is also a Hirzebruch surface.} Our task therefore reduces to
determining how these surfaces intersect one another.

\begin{figure}
\begin{center}
\includegraphics[scale=1.0]{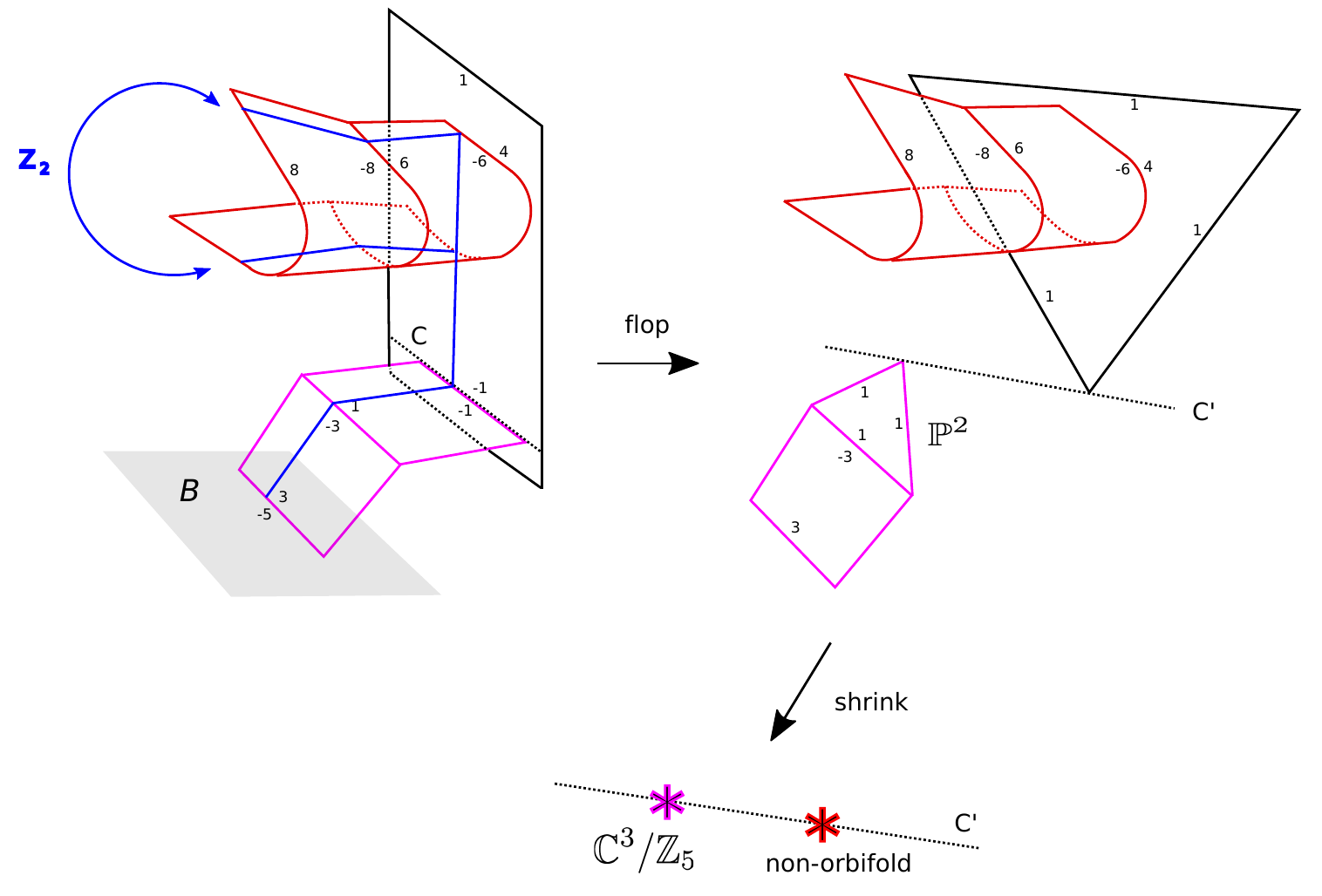}
\end{center}
\caption{Geometry of the $-5$ theory. In blue we have drawn the $E_6$ fiber over the generic point of the ruling, the $\mathbb{F}_6$ surface however meets the $\mathbb{F}_1$ corresponding to the center of the affine $E_6$ Dynkin diagram along a double section, which gives rise to the monodromy corresponding to the $\mathbb{Z}_2$ outer automorphism projecting $E_6$ to $F_4$.
In the 5D limit, the curve $C'$ decompactifies. As it is similar to other examples already presented, we have omitted
the other 5D SCFT limit described by pure $F_4$ gauge theory.}\label{minus5}
\end{figure}

The key difference from the cases with a simply laced algebra is the presence
of monodromy. So, starting from the affine Dynkin diagram for $e_{6}$, we see
that we now have only five surfaces, which intersect as:%
\begin{equation}
n=5:\begin{gathered}
\xymatrix{ &  & *++[o][F-]{\mathbb{F}_{a_5}}\ar@{-}[d]\\
&  &*++[o][F-]{ \mathbb{F}_{a_4}}\ar@{<=}[d]\\
*++[o][F-]{\mathbb{F}_{a_1}}\ar@{-}[r] & *++[o][F-]{\mathbb{F}_{a_2}} \ar@{-}[r]&*++[o][F-]{\mathbb{F}_{a_3}}}
\end{gathered} \label{nfive}%
\end{equation}
where here, we assume that the $\mathbb{Z}_{2}$ outer automorphism acts as a
reflection along the vertical axis of the affine $\mathfrak{e}_{6}$ Dynkin diagram,
yielding the affine $\mathfrak{f}_4$ Dynkin diagram as shown above.
Following the same reasoning used previously, we therefore conclude that
$a_{1}=3$, $a_{2}=1$ and $a_{3}=1$. The intersection of these surfaces follows
the same pattern outlined in the simply laced case. Now, to determine the
degree of the Hirzberuch surface $\mathbb{F}_{a_{4}}$, we observe that the
surface $\mathbb{F}_{a_{3}}=\mathbb{F}_{1}$ which it intersects can also be
viewed as a $\mathbb{P}^{2}$ blown up at one point. Owing to the monodromy in
the fiber, we see that this intersection locus must be a $\mathbb{P}^{1}$, and
must also provide a double cover of the hyperplane class $H$ of this
$\mathbb{P}^{2}$, and must also not intersect the exceptional divisor coming
from the blowup. This uniquely fixes the divisor class $C$ inside the
$\mathbb{P}^{2}$ to be $2H$, i.e., the vanishing locus of a homogeneous degree
two polynomial. The self-intersection of $C$ in
the $\mathbb{P}^{2}$ is:%
\begin{equation}
C\cdot_{\mathbb{P}^{2}}C=4,
\end{equation}
so the local geometry in the Calabi--Yau is $\mathcal{O}(4)+\mathcal{O}%
(-6)\rightarrow\mathbb{P}^{1}$. From this, we conclude that $a_{4}=6$.
Proceeding up in the vertical directions of line (\ref{nfive}), there are no
further effects from monodromy, and we find $a_{5}=8$. Summarizing, then, the
configuration of Hirzebruch surfaces is:%
\begin{equation}
n=5:
\begin{gathered}
\xymatrix{ &  & *++[o][F-]{\mathbb{F}_{8}}\ar@{-}[d]\\
&  &*++[o][F-]{ \mathbb{F}_{6}}\ar@{<=}[d]\\
*++[o][F-]{\mathbb{F}_{3}}\ar@{-}[r] & *++[o][F-]{\mathbb{F}_{1}} \ar@{-}[r]&*++[o][F-]{\mathbb{F}_{1}}}
\end{gathered}
\end{equation}
Note that the double arrow in the Dynkin diagram indicates that $\mathbb{F}_1$ and $\mathbb{F}_6$ meet along
a bisection of the ruling on $\mathbb{F}_1$.

In Figure \ref{minus5} we illustrate how a flop is needed to proceed
to the canonical 5D fixed point.

\subsection{$n=7$ Theory}

Finally, consider the case of the $-7$ curve theory. This case is different
from the previous ones because it contains matter fields in the 6D theory. We
realize an $\mathfrak{e}_{7}$ gauge theory with a half hypermultiplet in the
$\mathbf{56}$, i.e., the fundamental representation. The Weierstrass model for
this geometry is (see e.g. \cite{Bershadsky:1996nh}):
\begin{equation}
y^{2}=x^{3}+st^{3}x+t^{5}.
\end{equation}
To determine the configuration of surfaces in the resolved geometry,
consider again the case of the $-8$ curve theory. In both
this and the $-7$ curve theory, the fiber at a generic point of the base
$P^{1}$ is a $II^{\ast}$ fiber. The collection of surfaces in the $-8$ curve
case is:%
\begin{equation}
n=8:\begin{gathered}
\xymatrix{
&  &  & *++[o][F-]{\mathbb{F}_{2}}\ar@{-}[d] &  &  & \\
*++[o][F-]{\mathbb{F}_{6}}\ar@{-}[r] & *++[o][F-]{\mathbb{F}_{4}}\ar@{-}[r] & *++[o][F-]{\mathbb{F}_{2}} \ar@{-}[r]& *++[o][F-]{\mathbb{F}_{0}} \ar@{-}[r]&
*++[o][F-]{\mathbb{F}_{2}} \ar@{-}[r]& *++[o][F-]{\mathbb{F}_{4}}\ar@{-}[r] & *++[o][F-]{\mathbb{F}_{6}}%
}
\end{gathered}\label{neightagain}
\end{equation}

\begin{figure}
\begin{center}
\begin{tabular}{cc}
\includegraphics[scale=1.1]{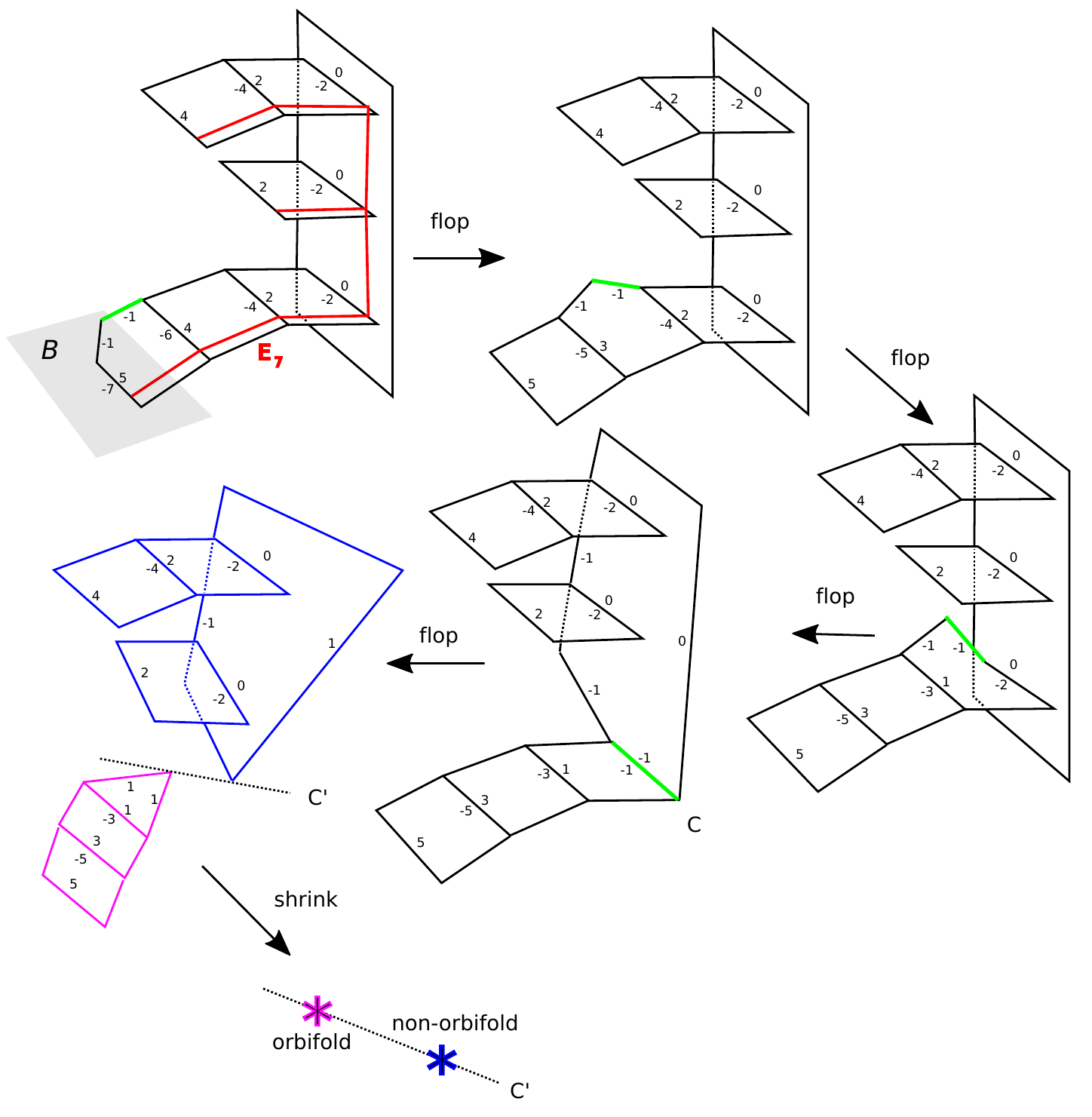}\\
\end{tabular}
\end{center}
\caption{
Geometry of the $-7$ theory. In red we have indicated the generic $E_7$ fiber along the ruling. We show explicitly the sequence of flop transitions leading to the 5D SCFT. In green we have indicate the curve that is being flopped at each step.
In the 5D limit, the curve $C'$ decompactifies. As it is similar to other examples already presented, we have omitted
the other 5D SCFT limit described by $E_7$ gauge theory with a half hypermultiplet in the
fundamental representation.}\label{minus7}
\end{figure}

Now, the only difference from the $n=8$ case is the presence of an additional
$\mathbb{P}^{1}$ in the degenerating fiber at the locus $s=0$. Based on this,
we can already deduce the general form of the configuration of surfaces:%
\begin{equation}
n=7:\begin{gathered}
\xymatrix{
&  &  & *++[o][F-]{\mathbb{F}_{1} }\ar@{-}[d]&  &  & \\
*++[o][F-]{S} \ar@{-}[r]& *++[o][F-]{\mathbb{F}_{3}} \ar@{-}[r]& *++[o][F-]{\mathbb{F}_{1}}\ar@{-}[r] & *++[o][F-]{\mathbb{F}_{1} }\ar@{-}[r]& *++[o][F-]{\mathbb{F}_{1}} \ar@{-}[r]&
*++[o][F-]{\mathbb{F}_{3} }\ar@{-}[r]&*++[o][F-]{ \mathbb{F}_{5}}
}
\end{gathered}
\end{equation}
where $S$ is a surface which intersects $\mathbb{F}_{3}$ along a
$\mathbb{P}^{1}$ of self-intersection $-3$ in the $\mathbb{F}_3$.
Now, to pass from the $n=8$ case to the $n=7$ case, we see
that we simply need to blowup a point on the $+6$ curve of the leftmost
$\mathbb{F}_{6}$ in line (\ref{neightagain}). After performing this blowup the
self-intersection of the curve shifts to $+5$, as one would expect for an
$\mathbb{F}_{5}$ surface. So, we denote this one point blowup of
$\mathbb{F}_{6}$ as $Bl^{(1)}\mathbb{F}_{6}$.  Summarizing, then, the
configuration of surfaces appearing for the $-7$ curve theory is:%
\begin{equation}
n=7:\begin{gathered}
\xymatrix{
&  &  & *++[o][F-]{\mathbb{F}_{1} }\ar@{-}[d]&  &  & \\
*++[o][F-]{Bl^{(1)}\mathbb{F}_{6}} \ar@{-}[r]& *++[o][F-]{\mathbb{F}_{3}} \ar@{-}[r]& *++[o][F-]{\mathbb{F}_{1}}\ar@{-}[r] & *++[o][F-]{\mathbb{F}_{1} }\ar@{-}[r]& *++[o][F-]{\mathbb{F}_{1}} \ar@{-}[r]&
*++[o][F-]{\mathbb{F}_{3} }\ar@{-}[r]&*++[o][F-]{ \mathbb{F}_{5}}
}
\end{gathered}
\label{nseven}%
\end{equation}
as shown in the upper left of  Figure \ref{minus7}. By the same token, further blowups on $\mathbb{F}_{6}$ lead us to $\mathfrak{e}_{7}$
gauge theories with additional half hypermultiplets. Similar considerations
also apply for the resolved geometries associated with fiber enhancements of
the other single curve theories.

Passing to the phase containing the canonical 5D fixed point is quite
tricky in this example.  As shown in Figure \ref{minus7}, a sequence
of flops must be performed until finally the resulting surfaces can
be contracted to two fixed points.

\newpage

\bibliographystyle{utphys}
\bibliography{5Dscft}

\end{document}